\documentclass[a4paper,11pt]{article}

\pdfoutput=1 
\usepackage{jheppub} 

\usepackage[utf8]{inputenc}
\usepackage{bbm}
\usepackage{multirow}

\usepackage{float}
\usepackage{graphicx}
\usepackage{color}
\usepackage{diagbox}

\definecolor{Red}{rgb}{1,0,0}
\definecolor{Green}{rgb}{0,0.7,0}
\definecolor{Blue}{rgb}{0,0,1}
\definecolor{Brown}{rgb}{0.392,0.165,0.165}
\definecolor{Black}{gray}{0}
\definecolor{Gray}{gray}{0.75}
\definecolor{clear}{gray}{1}

\newcommand{\mr}{\mathrm}
\renewcommand{\arraystretch}{1.0}
\newcommand{\refeq}[1]{(\ref{#1})}

\newcommand\sfrac[2]{{\textstyle \frac{#1}{#2}}}

\newcommand\comment[1]{}

\title{Seesaw neutrinos with one right-handed singlet field 
and a second Higgs doublet}
\author{D. Jur\v{c}iukonis,}
\author{T. Gajdosik,}
\author{and A. Juodagalvis}

\affiliation{Vilnius University, Institute of Theoretical Physics and Astronomy, \\
  Saul\.{e}tekio av.\ 3, Vilnius 10257, Lithuania}

\emailAdd{darius.jurciukonis@tfai.vu.lt}
\emailAdd{thomas.gajdosik@cern.ch}
\emailAdd{andrius.juodagalvis@tfai.vu.lt}

\abstract{
We study parameters of an extension of the Standard Model. 
The neutrino sector is enlarged by one right-handed singlet 
field, allowing for the seesaw mechanism type-I, and the Higgs
sector contains one additional doublet,
which contributes to light neutrino masses through
one-loop radiative corrections.
Employing an approximation for the effective light neutrino 
mass matrix we express the masses of the light neutrinos 
analytically, allowing us to parameterize the Yukawa couplings 
to neutrinos by the experimental measurements on the neutrino 
sector and only two free parameters. 
We focus on a CP-conserving Higgs potential 
for which we present the allowed ranges of the input parameters
and a statistical overview over the possible values of the 
Yukawa couplings. 
}

\begin{document} 
\maketitle
\flushbottom

\section{Introduction}

The precise interpretation of the neutral lepton fields in the particle
physics Lagrangian is not settled yet, owing to the very small mass of
the known neutrinos and the weakness of their interaction with other
particles~\cite{PDG2018}.
The observed neutrino oscillations support the notion that
neutrinos have non-vanishing masses, calling for a modification of the
Standard Model (SM).
The size of the neutrino mass is not the only puzzle to solve.
Absence of an electrical charge allows neutrinos to be
their own antiparticles. The nature of the neutrinos -- whether they
are Dirac or Majorana particles -- might be determined by future
experiments~\cite{Dolinski:2019nrj}. For the experimental constraints 
see~\cite{Anton:2019wmi, KamLAND-Zen:2016pfg, Aker:2019uuj}.

The Standard Model considers neutrinos as massless.
Adding heavy right-handed neutral
singlets and additional Higgs doublets, the authors of
ref.~\cite{Grimus:1989pu} combined the seesaw mechanism (type-I)
with the radiative mass generation. The spontaneous
symmetry breaking of the SM gauge group leads to a
Dirac mass term for neutrinos. The assumption that neutrinos are
Majorana particles allows an additional term in the Lagrangian,
namely, the Majorana mass term for the heavy singlets.

The model parameters allow small masses of the light
neutrinos that are compatible with the experimental observations. We
use this model in the formulation of Grimus and
Lavoura~\cite{Grimus:2002nk,Grimus:2002prd}, restricting the number of
additional Higgs doublets to one.
The case of three additional heavy
neutrino fields was studied e.g.\ in
refs.~\cite{AristizabalSierra:2011mn,Dev:2012sg}.  We assume
only one heavy neutrino field and consider only 1-loop corrections
to the neutrino mass matrix.
Ibarra and Simonetto~\cite{Ibarra:2011gn} analysed this scenario 
in the decoupling limit by renormalization group methods and 
predicted qualitatively our quantitative results. 
Our preliminary results were presented at several
conferences~\cite{Jurciukonis:2012jz,Jurciukonis:2012ft,
  Gajdosik:2013gpa,Jurciukonis:2014sma,Gajdosik:2015jja}.
This paper provides a
more complete description of the performed numerical analysis.
We reduce the number of free model parameters by linking the model
predictions with experimental neutrino observables.

Our extended model has several subsets of parameters. The neutrino
sector is characterized by the mass of the heavy neutrino 
and the strength of the coupling to the neutral Higgs fields.
The masses of the three light neutrinos are the result of our model
parameters. They are subject to experimental constraints,
namely the experimental neutrino mass differences, $\Delta m^2_{21}$
and $\left|\Delta m_{31}^{2}\right|$,
as well as the experimental neutrino oscillation
angles $\theta_{12}$, $\theta_{13}$, and
$\theta_{23}$~\cite{deSalas:2017kay}.
We follow the ideas from ref.~\cite{Xing:2011ur} on
neutrino oscillation angle estimation from the neutrino mixing matrix.
More details are given in
appendix~\ref{Appendix-oscillation-angles}.
It should be noted that experimental data is
usually interpreted in the ``$3 \times 3$'' neutrino mixing
model~\cite{PDG2018,deSalas:2017kay},
i.e.\ three flavoured neutrinos are considered
as mixed states of three neutrino mass eigenstates. We do not
attempt to reinterpret the experimental results in the context of an
extended neutrino model. 

We parameterize the Higgs sector following the analysis
of Haber and O'Neil~\cite{Haber:2010bw}.
The Yukawa couplings are parameterized similarly to Grimus and
Lavoura~\cite{Grimus:2002nk,Grimus:2002prd}, which coincide
with \cite{Haber:2010bw} in the Higgs sector.
For the numerical analysis we take the mass of
the SM-like Higgs boson as $m_{h}=125.18$~GeV~\cite{PDG2018}
and allow the masses of two other neutral Higgs bosons
to vary in the range from $m_{h}$ to 3000~GeV.

Using cosmological arguments the PLANCK collaboration
finds~\cite{Aghanim:2018eyx} that the sum of all light neutrino
masses is limited by $\sum m_{\nu} < 0.12$~eV.
The earlier upper bound estimate was notably larger:
$\sum m_{\nu}<0.23$~eV \cite{Ade:Planck}.
If the new bound is correct,
the overall scale of the neutrino masses must be smaller,
and the mass of the lightest neutrino could be much smaller 
than the masses of the other neutrinos, especially for the 
inverted hierarchy. 
As a matter of fact, the lightest neutrino has no mass in the model of
ref.~\cite{Grimus:1989pu} with only one heavy neutrino. We call this 
setup the Grimus-Neufeld model.
However, this Grimus-Neufeld model is compatible 
with the results of \cite{Emami,Aghanim:2018eyx} and 
is fully consistent with the current experimental neutrino data.

The outline of the paper is the following. Section \ref{framework} reviews
the seesaw mechanism and the formalism of the two-Higgs-doublet model
as it is used in our analysis. 
Section \ref{framework:improved couplings} shows the analytic determination
of the neutrino masses that can be used to replace free model parameters
by the measured neutrino mass differences and mixing angles.
Section \ref{NumRes} describes our main results, namely, the analysis 
of the free model parameters and restrictions for the Higgs sector. 
Our findings are summarized in
section \ref{summary}. For completeness, appendix
\ref{Appendix-b-vectors} describes the features of the weight
vectors $b_i$ that relate the scalar Higgs fields to their mass
eigenfields, appendix
\ref{Appendix-oscillation-angles} gives the details of the oscillation
angle calculation, and appendix \ref{Appendix-2HDM} summarizes the restrictions
that we apply to the parameters of the 2HDM potential.

\section{Description of the model}
\label{framework}

We discuss an extension of the Standard Model with enlarged Higgs
and neutrino sectors. Our main interest is the neutrino sector.
Since we need the Higgs sector for the radiative neutrino masses,
we give a short overview of the properties of the Higgs sector
that we use in our calculations.

\subsection{The Higgs sector}
\label{framework:Higgs}
The authors of ref.~\cite{Haber:2006ue} discuss the basis independent
formulation of the general two-Higgs-doublet model (2HDM).
Using their definition of the Higgs basis, we can write the
two complex doublets of our model in a unique way
\begin{equation}\label{higgsbasis}
\phi_{1}
=
\left(
\begin{array}{c} G^{+} \\ \frac{1}{\sqrt{2}}
( v + \mathcal{H}^{0}_{1r} + i G^{0} )  \end{array}
\right)\ ,
\qquad
\phi_{2}
=
\left(
\begin{array}{c} \mathcal{H}^{+} \\ \frac{1}{\sqrt{2}}
( \mathcal{H}^0_{2r} + i \mathcal{H}^0_{2i} )  \end{array}
\right)
\enspace ,
\end{equation}
where the vacuum expectation value (VEV) $v \simeq 246$~GeV and the
Goldstone bosons $G^{0}$ and $G^{+}$ appear only in the first
Higgs doublet $\phi_{1}$. The tree-level relations between the 
basis independent parameters defining the Higgs potential 
and the parameters describing
the physical states are linear and can be easily inverted. This
feature allows us to use the VEV, the masses of the physical Higgs
bosons, $m_{H_1^0}$, $m_{H_2^0}$, $m_{H_3^0}$, and $m_{H^+}$,
and their mixing angles $\vartheta_{12}$ and $\vartheta_{13}$ as input
parameters.

The mass eigenstate for the charged Higgs boson corresponds directly
to the field $\mathcal{H}^{+}$ with the mass $m_{H^+}$, but the
mass eigenstates for the neutral Higgs bosons with the masses
$m_{H_1^0}$, $m_{H_2^0}$, and $m_{H_3^0}$, respectively, are
linear superpositions of the neutral fields $\mathcal{H}^{0}_{1r}$,
$\mathcal{H}^0_{2r}$, and $\mathcal{H}^0_{2i}$. Following the
formulation of Grimus and Lavoura~\cite{Grimus:2002nk,Grimus:2002prd}
these linear superpositions are conveniently expressed by
\begin{equation}
\label{sumb}
H_k^0
 =
\phi_{b_k}^0
 =
\sqrt{2} \, \mathrm{Re}(b_k^{\dagger} \overline{\phi}^0)
 =
\sqrt{2} \sum_{j=1}^{n_H} \mathrm{Re}(b_{kj}^* \overline{\phi}_j^0)
 =
 \frac{1}{\sqrt{2}} \sum_{j=1}^{n_{H}}
 \left( b^{*}_{kj} \overline{\phi}^{0}_j
      + b_{kj} \overline{\phi}^{0\,*}_j \right)
\enspace ,
\end{equation}
where $\overline{\phi}^{0}$ are the neutral parts of the Higgs doublets
without the VEV: $\overline{\phi}_{1}^{0} = \phi_{1}^{0} - v/\sqrt{2}$
and $\overline{\phi}_{2}^{0} = \phi_{2}^{0}$.
There are $2n_H$ unit-length ``$b$-vectors'' ($b_k \in
\mathbbm{C}^{n_H}$) of dimensions $n_H \times 1$,
where $n_H$ is the number of Higgs doublets, i.e.\ $n_H=2$ in the 2HDM.
We discuss those
vectors in the general case in appendix~\ref{Appendix-b-vectors}.
There we also show how
to obtain the following parametric values for the vectors $b_k$:
\begin{equation}
\label{bVectSetTh}
b_{G^0} = \left( \begin{array}{c} i \\ 0 \end{array} \right)
, \hspace{0.2cm}
b_1 = \left( \begin{array}{c}
 \mathrm{c}_{12} \mathrm{c}_{13}
 \\ -\mathrm{s}_{12} - i \mathrm{c}_{12} \mathrm{s}_{13}
\end{array} \right)
, \hspace{0.2cm}
b_2 = \left( \begin{array}{c}
 \mathrm{s}_{12} \mathrm{c}_{13}
 \\ \mathrm{c}_{12} - i \mathrm{s}_{12} \mathrm{s}_{13}
\end{array} \right)
, \hspace{0.2cm}
b_3 = \left( \begin{array}{c}
 \mathrm{s}_{13} \\
 i \mathrm{c}_{13}
\end{array} \right)
\enspace ,
\end{equation}
where $\mathrm{c}_{1j} = \cos\vartheta_{1j}$ and
$\mathrm{s}_{1j} = \sin\vartheta_{1j}$ ($j=2, 3$) are determined by the angles
$\vartheta_{1j}$ that describe the mixing of the neutral Higgs fields.

Restricting ourselves to the CP conserving case we use the analysis
of ref.~\cite{Haber:2010bw}, where the authors discuss
the CP-invariant Higgs potential in the 2HDM framework
under various basis-independent conditions.
The possible overall phase, that can be written in front of
the second Higgs doublet and that acts like a mixing angle
$\vartheta_{23}$ between $\mathcal{H}^0_{2r}$ and $\mathcal{H}^0_{2i}$,
is used to define the CP-property of the mass eigenstates,
corresponding to their coupling to gauge bosons. The choice is
$H^{0}$ to be CP-even and $A^{0}$ to be CP-odd.

This assignment does not order the masses of the neutral Higgs bosons:
both $m_{H} < m_{A}$ and $m_{H} > m_{A}$ are possible, giving us 
two conditions (case~I and case~II), which are listed in
table~\ref{table1}.
We still assume the fixed SM-like Higgs mass $m_{H^0_1} \equiv m_{h}$ 
to be smaller than the other two: $m_{h} < m_{H,A}$. 
Authors of \cite{Haber:2006ue} argue that one can assume
$-\frac{\pi}{2} \leqslant \vartheta_{12},\vartheta_{13} < \frac{\pi}{2}$
without the loss of generality.
We perform the numerical analysis of the neutrino
mass spectrum considering the named two cases, but using only the 
single mixing angle $(\beta-\alpha)$.

\def\BigColSep{\setlength{\arraycolsep}{0pt}}
\renewcommand{\arraystretch}{1.2}
\begin{table*}
\begin{center}
\begingroup\BigColSep
\begin{tabular}{|c|c|c|}
\hline \hline
 & I & II \\
\hline
 & $\vartheta_{13} = 0$ & $\vartheta_{12} = 0$
\\
\hline
 & $m_{H}<m_{A}$ & $m_{H}>m_{A}$ 
\\
\hline
$b_1$
& $\left( \begin{array}{c} \mathrm{c}_{12} \\
  - \mathrm{s}_{12} \end{array} \right) \equiv 
  \left( \begin{array}{c} \mathrm{s}_{\beta-\alpha} \\
  - \varepsilon\mathrm{c}_{\beta-\alpha} \end{array} \right)$
& $\left( \begin{array}{c} \mathrm{c}_{13} \\
  -i \mathrm{s}_{13} \end{array} \right) \equiv 
  \left( \begin{array}{c} \mathrm{s}_{\beta-\alpha} \\
  i \varepsilon\mathrm{c}_{\beta-\alpha} \end{array} \right)$
\\
\hline
$b_2$
& $\left( \begin{array}{c} \mathrm{s}_{12} \\
  \mathrm{c}_{12} \end{array} \right) \equiv 
  \left( \begin{array}{c} \varepsilon\mathrm{c}_{\beta-\alpha} \\
  \mathrm{s}_{\beta-\alpha} \end{array} \right)$
& $\left( \begin{array}{c} 0 \\ 1 \end{array} \right)$
\\
\hline
$b_3$
& $\left( \begin{array}{c} 0 \\ i \end{array} \right)$
& $\left( \begin{array}{c} \mathrm{s}_{13} \\
  i \mathrm{c}_{13} \end{array} \right) \equiv 
  \left( \begin{array}{c} - \varepsilon\mathrm{c}_{\beta-\alpha} \\
  i\mathrm{s}_{\beta-\alpha} \end{array} \right)$
\\
\hline\hline
\end{tabular}
\endgroup
\end{center}
\caption{Basis-independent conditions for a CP-conserving 2HDM 
scalar potential and vacuum~\cite{Haber:2010bw}. $\vartheta_{ij}$ are 
the mixing angles of the neutral Higgses and $\beta-\alpha$ is the 
invariant angle constructed from the angle $\alpha$ which mixes the 
CP-even Higgs bosons and the angle $\beta$ which relates the values of 
the VEV's; $\varepsilon \equiv \mathrm{sgn}(\beta-\alpha)$ is a 
pseudo-invariant quantity; 
$m_{H}$ and $m_{A}$ denote the masses for the CP-even and CP-odd Higgses. 
Relations between neutral Higgs fields and angular factors are explained in 
more detail in appendix~\ref{Appendix-2HDM} and in ref.~\cite{Haber:2006ue}. 
Our case~I corresponds
 to the case~I of~\cite{Haber:2010bw}, whereas our case~II corresponds
 to the case~IIa of~\cite{Haber:2010bw}.} \label{table1}
\end{table*}
\renewcommand{\arraystretch}{1.0}

\subsection{The Yukawa couplings}
\label{framework:Yukawa}
Using the
vector-and-matrix notation, the Yukawa Lagrangian for the leptons is
expressed~\cite{Grimus:2002nk,Grimus:2002prd} as
\begin{equation}
\label{Yukawa}
\mathcal{L}_\mathrm{Y} = - \sum_{k=1}^{n_H=2}\,
\left( \phi_k^\dagger \bar \ell_R \Gamma_k
+ \tilde \phi_k^\dagger \bar \nu_R \Delta_k \right)
    \left(\begin{array}{c}
      \nu_L \\
      \ell_L
    \end{array}\right)
+ \mathrm{H.c.} ,
\end{equation}
where $\tilde \phi_k = i \tau_2 \phi_k^\ast$. The quantities
$\ell_R$ and $\nu_R$ are the vectors of the
right-handed charged leptons and the right-handed projection of
the neutrino singlets, respectively.
$\ell_L$ and $\nu_L$ form the lepton doublet under the
weak interactions
and combine with the Higgs doublets $\phi_{k}$ to form
$SU(2)_{\mathrm{weak}}$-invariant terms.
They are also vectors in the generation space of dimension
$n_L = 3$.
The Yukawa
coupling matrices $\Gamma_k$ have the dimension $n_L \times n_L$, while
$\Delta_k$ have the dimension $n_R \times n_L$, where $n_R$ is
the number of the singlet neutrino fields, $n_R = 1$ in our case.

Taking the bilinear terms of eq.~(\ref{Yukawa}), which means taking
only the VEV from the Higgs doublets, we get the Dirac mass terms
for charged leptons and neutrinos, assuming the charged leptons
to be in their mass eigenstates:
\begin{equation}\label{M_ell}
M_\ell = \frac{v}{\sqrt{2}}\, \Gamma_1
\enspace\doteq
\mathrm{diag} \left( m_e, m_\mu, m_\tau \right)
\end{equation}
and
\begin{equation}\label{M_D}
 M_D = \frac{v}{\sqrt{2}} \Delta_1
\enspace .
\end{equation}
These matrices
have to be diagonalized using the singular-value decomposition (SVD)
like in the SM to get the correct definition for the mass eigenstates
that will describe the physical particles. Having done this
transformation to the mass eigenstates, which we write down as the
fields appearing in eq.~(\ref{Yukawa}), the respective transformation
matrices reappear in two unique combinations, $V_{\mathrm{CKM}}$
and $V_{\mathrm{PMNS}}$, in the interactions
with the charged gauge bosons $W^{\mp}$ or the charged scalar bosons
$H^{+}$ and $G^{+}$, giving the charged current Lagrangian
\begin{equation}
\label{PMNS}
\mathcal{L}_{\mathrm{cc}}
= \frac{g}{\sqrt{2}}\, W_{\mu}^{-}
  \bar{\ell}_{L} \gamma^{\mu} P_L \nu_{L}
+ \mathrm{H.c.}
= \frac{g}{\sqrt{2}}\, W_{\mu}^{-}
  \bar{\ell}_{L} \gamma^{\mu} P_L V_{\mathrm{PMNS}} \, \zeta 
+ \mathrm{H.c.}
\enspace ,
\end{equation}
where $g$ is the $SU(2)$ gauge coupling constant and $\zeta$ 
stands for the neutrino mass eigenstates. We give this part of
the Lagrangian only as a reference, to show what neutrino experiments
measure, as this PMNS matrix $V_{\mathrm{PMNS}}$ is the basis for the
interpretation of experimental data in the ``$3\times3$'' neutrino mixing
model~\cite{PDG2018}.

\subsection{Neutrinos at tree level}
\label{framework:neutrinos}
The singlet neutrinos, added to the SM, are neutral with respect to all
gauge groups of the SM. This offers the possibility that they are
Majorana particles, allowing to write a Majorana mass term for them.
Since the Lagrangian has to be a scalar with respect to Lorentz
transformations, we have to combine a spinor with itself in a Lorentz
invariant way. The Dirac spinors can only be combined using the charge
conjugation matrix $\mathbf{C}$, which also appears in the definition
of the Lorentz covariant conjugation\footnote{A very clear and
exhaustive description of the difference between Majorana and Dirac
spinors is given in ref.~\cite{Pal:2010ih}.}
\begin{equation}\label{lcc}
\hat{\Psi}
:=
\gamma^{0} \mathbf{C} \Psi^{*}
=
- \mathbf{C} \bar{\Psi}^{\top}
\enspace ,
\end{equation}
where $\Psi$ is a Dirac spinor.
The Majorana condition can now be written as
\begin{equation}\label{majorana-condition}
\hat{\Psi}_{M}
=
\eta_{\Psi} \Psi_{M}
\enspace ,
\end{equation}
where $\eta_{\Psi}$ is the Majorana phase. Assuming $\nu_{R}$ to
be $n_{R}$ Majorana fermions we can write down the Majorana mass term as
\begin{equation}\label{majorana-mass-term}
\mathcal{L}_{\mathrm{Majorana\text{-}mass}}
=
- \sfrac{1}{2} \bar{\nu}_{R} M_{R} \hat{\nu}_{R} + H.c.
=
  \sfrac{1}{2} \bar{\nu}_{R} M_{R} \mathbf{C} \bar{\nu}_{R}^{\top}  + H.c.
\enspace ,
\end{equation}
where the order of $M_{R}$ and $\mathbf{C}$ is irrelevant, as these
matrices act on different indices of the spinor $\nu_{R}$: $\mathbf{C}$
is a $4\times4$ matrix, connecting the spinor indices of $\nu_{R}$,
whereas $M_{R}$ is a symmetric $n_{R}\times n_{R}$ matrix, acting on the
``generation'' index of $\nu_{R}$.
Since in our case $\nu_{R}=1$, the Majorana mass matrix of the
heavy singlet $M_{R}$ is just a number.

The mass terms for the neutrinos, including the Dirac mass terms
originating from the Yukawa terms in eq.~(\ref{Yukawa}), can be written as
\begin{eqnarray}\label{nu-mass}
  \mathcal{L}_{\nu\text{-}\mathrm{mass}}
&=&
- \bar{\nu}_{R} M_{D} \nu_{L}
- \sfrac{1}{2} \bar{\nu}_{R} M_{R} \hat{\nu}_{R}
+ H.c.
\nonumber\\
&=&
- \sfrac{1}{2} \bar{\nu}_{R} M_{D} \nu_{L}
- \sfrac{1}{2} \bar{\hat{\nu}}_{L} M_{D}^{\top} \hat{\nu}_{R}
+ \sfrac{1}{2} \bar{\nu}_{R} M_{R} \mathbf{C} \bar{\nu}_{R}^{\top}
+ H.c.
\nonumber\\
&=&
- \sfrac{1}{2}
\left(\begin{array}{cc}
  \bar{\hat{\nu}}_{L} & \bar{\nu}_{R}
\end{array}\right)
\left(\begin{array}{cc}
  M_{L} & M_{D}^{\top} \\
  M_{D} & M_{R}
\end{array}\right)
\left(\begin{array}{c}
  \nu_{L}
 \\ \hat{\nu}_{R}
\end{array}\right)
+ H.c.
\end{eqnarray}
and can be written in a compact form by introducing the
$(n_L+n_R) \times (n_L+n_R)$ symmetric neutrino mass matrix
\begin{equation}
\label{Mneutr}
\renewcommand{\arraystretch}{0.8}
M_{\nu} =
\left(\begin{array}{cc} 0 & M_D^{\top} \\
M_D & M_R \end{array} \right)
\enspace .
\end{equation}
The Majorana mass matrix of the light neutrinos is vanishing at tree
level, $M_L=0$.

The neutrino mass matrix $M_{\nu}$ can be
diagonalized~\cite{Grimus:1989pu,Grimus:2002nk,Grimus:2002prd} 
using the properties
of the singular-value decomposition of a symmetric matrix, 
or Takagi factorization~\cite{Hahn:2006hr} 
\begin{equation}
\label{Mtotal}
U^{\top} M_{\nu}\, U = \hat m
= \mathrm{diag}
\left( m_{1}, m_{2}, m_{3}, m_{4} \right),
\end{equation}
where $m_{i}$ are real and non-negative. Following the conventions 
of~\cite{deSalas:2017kay} we adopt the mass-ordering 
$m_{1} \le m_{2} < m_{3} \ll m_{4}$ for the normal hierarchy and 
$m_{3} \le m_{1} < m_{2} \ll m_{4}$ for the inverted hierarchy of 
the neutrino mass spectrum.
In order to implement the seesaw mechanism
\cite{GellMann:1980vs,Schechter:1980gr} we assume that the elements of
$M_D$ are of order $m_D$ with
$m_D \ll M_R$. Then, the neutrino masses $m_{i}$ with
$i=1,\ldots,n_L$ (where $n_L=3$), are of order $m_D^2/M_R$, while the
mass $m_{4}$ is of order $M_R$.

At tree-level, $\hat{m}$ contains only two non-vanishing neutrino
masses: the
mass $m_{4}^{\text{tree}}$ of the heavy neutrino
$\zeta_{4}^{\text{tree}}$ and the mass of
one light neutrino that is generated by the seesaw mechanism.
We will refer to it as the ``seesaw neutrino'' $\zeta_{s}^{\text{tree}}$
with the mass $m_{s}^{\text{tree}}$.
(The neutrino states in the mass
basis are denoted as $\zeta$ to distinguish them from the flavour
eigenstates denoted as $\nu$.)
The
remaining two neutrino states are massless at tree-level. Since the
radiative corrections~\cite{Grimus:1989pu} generate only one mass,
one of these two states will stay massless. We call this state $\zeta_{o}$
with the mass $m_{o} = 0$. The seesaw neutrino $\zeta_{s}$ has
the mass $m_{s}$.
The remaining third light neutrino
$\zeta_{r}$ has the mass $m_{r}$. As argued in ref.~\cite{Grimus:1989pu},
the loop generated (i.e.\ radiative) mass $m_{r}$
can be of the same order as the seesaw
generated mass $m_{s}^{\text{tree}}$.
Hence we do not impose an ordering between these 
two states ($m_{s}$ and $m_{r}$).
Combining these two possibilities of the ordering with
the normal or inverted hierarchy we can have four arrangements of indices 
between the names $o$, $r$, and $s$, and the numbers $1$, $2$, and $3$,
as displayed in Table~\ref{index-table}. 
Since the formulation of the theoretical basis does not care about the
numbering, we stay with the names and refer to Table~\ref{index-table} 
only when implementing the physical values.
\begin{table*}
\begin{center}
\begin{tabular}{|c|c|c|c|}
\hline
\diagbox{scenario}{index}              & $o$ & $r$ & $s$
 \\[1pt] \hline \hline
& & & \\[-12pt] NH                     &  1  &  2  &  3 
 \\[1pt] \hline 
& & & \\[-12pt] $\overline{\text{NH}}$ &  1  &  3  &  2 
 \\[1pt] \hline \hline 
& & & \\[-12pt] IH                     &  3  &  1  &  2 
 \\[1pt] \hline 
& & & \\[-12pt] $\overline{\text{IH}}$ &  3  &  2  &  1 
 \\[1pt] \hline
\end{tabular}
\end{center}
\caption{Index arrangements between the naming and numbering of the 
light neutrino states. The overbarred scenarios describe the case, 
when the loop-generated mass $m_r$ becomes bigger than the loop-corrected
seesaw mass $m_s$. The mass $m_o$ is always $0$ in our model.}
\label{index-table}
\end{table*}

It is useful to
decompose the $(n_L+n_R) \times (n_L+n_R)$ unitary matrix $U$ 
from eq.~(\ref{Mtotal}) into two submatrices
\cite{Grimus:1989pu,Grimus:2002nk,Grimus:2002prd}
\begin{equation}\label{U}
U = \left( \begin{array}{c} U_L \\ U_R^\ast \end{array} \right),
\end{equation}
where the submatrix $U_L$ is of size $n_L \times (n_L+n_R)$ and the
submatrix $U_R$ is $n_R \times (n_L+n_R)$. These submatrices obey
certain unitarity relations:
\begin{equation}\label{uni}
U_L U_L^\dagger = \mathbbm{1}_{n_L} \,, \quad
U_R U_R^\dagger = \mathbbm{1}_{n_R} \,, \quad
U_L U_R^{\top} = 0_{n_L \times n_R} \,, \quad \mathrm{and} \quad
U_L^\dagger U_L + U_R^{\top} U_R^* = \mathbbm{1}_{n_L+n_R}
\enspace .
\end{equation}
Combining with eq.~(\ref{Mtotal}), we can obtain the following relations:
\begin{equation}\label{Urelat}
U_L^* \hat{m} U_L^\dagger = 0, \quad
U_R \hat{m} U_L^\dagger = M_D, \quad \mathrm{and} \quad
U_R \hat{m} U_R^{\top} = M_R
\enspace .
\end{equation}

With these submatrices of $U$,
the left- and right-handed neutrinos
can be written as linear superpositions
of the $n_L+n_R$ physical Majorana neutrino fields $\zeta_{\alpha}$
(to the remainder of this section, we omit the superscript
``$\text{tree}$''):
\begin{equation}\label{chi}
\nu_L = U_L P_L \zeta,
\quad \mathrm{and} \quad
\hat{\nu}_R = U_R^{*} P_L \zeta
\quad \mathrm{or} \quad
\nu_R = U_R P_R \zeta
\enspace ,
\end{equation}
where $P_L$ and $P_R$ are the projectors of chirality.

Switching to the physical Majorana mass states $\zeta$, we have to
express the field couplings using the matrices $U_{L}$ and $U_{R}$.
Neutrino interaction with the $Z$ boson is given by
\begin{equation}
\mathcal{L}_{\mathrm{nc}}^{(\nu)} = \frac{g}{4 c_w}\, Z_\mu
\bar{\zeta} \gamma^\mu \left[ P_L \left( U_L^\dagger U_L \right)
- P_R \left( U_L^{\top} U_L^\ast \right) \right] \zeta \, ,
\label{Zinter}
\end{equation}
where $c_w$ is the cosine of the Weinberg angle.
The Yukawa couplings for the neutral scalars
take the form
\begin{align}
\mathcal{L}_\mathrm{Y}^{(\nu)} \left( H_{k}^0 \right) = 
 -\frac{1}{2 \sqrt{2}}\,
 \sum\limits_{k=1}^{2n_H} H_{k}^0 \,
 \bar{\zeta} \Big[
& \left( U_R^\dagger \Delta_{b_k} U_L
+ U_L^{\top} \Delta_{b_k}^{\top} U_R^\ast \right) P_L \notag \\
+& \left( U_L^\dagger \Delta_{b_k}^\dagger U_R
+ U_R^{\top} \Delta_{b_k}^\ast U_L^\ast \right)P_R
\Big] \zeta \, ,
\label{neutralYuk}
\end{align}
where we treat the Goldstone boson $G^{0}$ as $H_{4}^{0}$.
The Yukawa coupling $\Delta_{b_k}$ is the result of rewriting
the Yukawa Lagrangian eq.~(\ref{Yukawa}) using the physical Higgs
fields defined in eq.~(\ref{sumb}):
\begin{equation}
\Delta_{b_k} = \sum_{j=1}^{n_{H}} (b_k)_{j} \Delta_{j} .
\end{equation}
The tree level quantities are used to calculate 1-loop corrections.

\subsection{Loop corrections to the neutrino masses}
\label{framework:loops}
We are interested in radiatively generated neutrino masses at one-loop
level~\cite{Grimus:2002nk}. The light neutrino Majorana mass term
$\delta M_L$ has the largest influence from the corrections to the
neutrino mass matrix, since this submatrix is zero at tree level,
$\left.M_L\right|_{\text{tree}}=0$. The contributions to the masses
from charge-changing currents are
subdominant~\cite{Grimus:2002nk,Grimus:2002prd,Pilaftsis:1991ug}.

Once the one-loop corrections are taken into account, the neutral
fermion mass matrix is given by~\cite{Grimus:2002nk}
\begin{equation}\label{M1}
M^{(1)}_\nu = \left( \begin{array}{cc}
\delta M_L & M_D^{\top}+\delta M_D^{\top} \\
M_D+\delta M_D  & \hat{M}_R+\delta M_R\end{array} \right)\approx
\left( \begin{array}{cc}
\delta M_L & M_D^{\top} \\
M_D  & \hat{M}_R\end{array} \right) \enspace .
\end{equation}
The one-loop
corrections to $\delta M_L$ originate via the self-energy functions
$\Sigma_L^{S(X)}(0)$ (where $X=Z,G^0,H_k^0$, $k=1,2,3$) that arise from the
self-energy Feynman diagrams. The contributions
$\Sigma_L^{S}(p^2)$ are evaluated at zero external
momentum squared ($p^2=0$). The neutrino couplings to
the $Z$, Higgs $H_k^0$ and Goldstone $G^0$ bosons
are determined by eqs.~(\ref{Zinter}) and (\ref{neutralYuk}). Each
diagram contains a divergent piece but the sum of the three
contributions yields a finite result. The expression for these one-loop
corrections is given by (see e.g.\ \cite{Grimus:2002nk})
\begin{eqnarray}
\delta M_L &=&
\sum_{k=1}^{3} \frac{1}{32 \pi^2}\, \Delta_{b_k}^{\top} U_R^\ast \hat m
\left(
  \frac{\hat {m}^2}{m_{H^0_k}^2}-\mathbbm{1}
 \right)^{-1}\hspace{-5pt}
 \ln\left(\frac{\hat {m}^2}{m_{H^0_k}^2}\right) U_R^\dagger \Delta_{b_k}
 \notag \\
&&+ \frac{3 g^2}{64 \pi^2 m_W^2}\, M_D^{\top} U_R^\ast \hat m
\left(
  \frac{\hat {m}^2}{m_Z^2}-\mathbbm{1}
 \right)^{-1}\hspace{-5pt}
 \ln\left(\frac{\hat {m}^2}{m_Z^2}\right) U_R^\dagger M_D
\enspace ,
\label{corrections}
\end{eqnarray}
where the sum index $k$ runs over all neutral physical Higgses $H^0_k$. 
The 1-loop corrections are defined in terms of tree level quantities.

\subsection{Parameters of the model}
\label{framework: parameters of the model}
As the Grimus-Neufeld model is a minimal extension of the Standard Model,
the only additions to the Lagrangian of the Standard Model are the 
heavy singlet Majorana mass term, eq.~(\ref{majorana-mass-term}), 
the Yukawa couplings to the heavy singlet fermion, $\Delta_{j}$, the
Yukawa couplings of the second Higgs doublet to the charged leptons, 
$\Gamma_{2}$, both given in eq.~(\ref{Yukawa}), and the Higgs potential
of the two Higgs doublets, that replaces the Higgs potential of the 
Standard Model. That gives us 
\begin{eqnarray}\label{parameterListLagrangian}
\{ 
 p_{i,\text{SM}} , p_{i,\text{2HDM}} , M_{R} , \Delta_{j} , \Gamma_{2}
\}
\end{eqnarray}
as the primary parameters of our model. 
$p_{i,\text{SM}}$ denotes the SM parameters like the masses of the charged
leptons or the Fermi coupling constant $G_{F}$. 
$p_{i,\text{2HDM}}$ stands for the parameterization of the 2HDM potential
and can be either the potential parameters $m_{ij}^{2}$ and $\lambda_{j}$ 
or, following the idea of~\cite{Grzadkowski:2018ohf,Grzadkowski:2019nwa}, 
the masses and physical couplings of the Higgs fields.
It means also that we assume the 
charged fermion fields to be in their mass eigenstates, making 
$\Gamma_{1} = \frac{\sqrt{2}}{v} \text{diag}[ m_{e} , m_{\mu} , m_{\tau} ]$ 
a diagonal matrix.

Following the guidelines of~\cite{Ofreid:2hdmwork2018,Ogreid:2018bjq} 
we can swap the parameters $p_{i,\text{2HDM}}$ for the masses
of the physical Higgs bosons, $m^{2}_{H_{i}^{0}}$ and $m^{2}_{H_{}^{\pm}}$, 
the physical couplings $e_{i}$ of the neutral Higgses $H_{i}^{0}$ 
to a pair of $W$-bosons, the selfcouplings 
$q_{i}$ of the neutral Higgses $H_{i}^{0}$ 
to a pair of charged Higgses, and the selfcoupling $q$ of the 
charged Higgses. 
But instead of using the 7 couplings $e_{i}$, $q_{i}$, and $q$, we just 
use the mixing angles of the neutral Higgses in the Higgs basis, as 
indicated by their use in the $b$-vectors, eq.~(\ref{bVectSetTh}), or
table~\ref{table1}.


\section{Reducing parameters by neutrino measurements}
\label{framework:improved couplings}

The main goal of this section is to show, how we can replace the 6 complex
parameters in $\Delta_{1}$ and $\Delta_{2}$ by the measured mass differences
of the light neutrinos, the entries of the PMNS matrix and two additional 
real parameters. Of course, this works only because not all of the 6 complex
parameters in $\Delta_{j}$ are physically independent.

Using the approximation to the contributions of the 1-loop corrections to the
neutrino mass matrix, eq.~(\ref{corrections}), 
we can relate the calculated neutrino masses 
to the measured neutrino mass differences.

Following~\cite{Grimus:2002nk} we treat only the effective $3\times 3$ 
light neutrino mass matrix $\mathcal{M}_{\nu}$, which is
a rank~1 matrix at tree level and equals
\begin{eqnarray}
\mathcal{M}_{\nu}^{\text{tree}} 
 = 
- M_D^{\top} M_{R}^{-1} M_D
\enspace .
\label{M3x3-tree}
\end{eqnarray}
Similarly to the treatment in \cite{Gajdosik:2015dvb}, we can write the
diagonalization of the tree-level neutrino mass matrix as
\begin{equation}
\label{Mtree-diag}
  V^{\top} \mathcal{M}_{\nu}^{\text{tree}}\, V
= 
- V^{\top} M_D^{\top} M_{R}^{-1} M_D V
= 
- \mathrm{diag}
  \left( 0 , 0 , m_{s}^{\text{tree}} \right),
\end{equation}
with the three column vectors $\vec{V}_{i}$ forming the unitary $3\times3$
matrix $V=({\vec V}_{o},{\vec V}_{r},{\vec V}_{s})$ and 
$m_{s}^{\text{tree}} > 0$.
This equation, eq.~(\ref{Mtree-diag}), leads to the
conditions for the vectors $\vec{V}_o$ and $\vec{V}_r$
\begin{equation}
\label{V-conditions-H1-0}
  M_{D} \cdot \vec{V}_{o}^{} 
=
  M_{D} \cdot \vec{V}_{r}^{} 
= 
  0 \enspace ,
\end{equation}
meaning that the neutrino states $\zeta_{o}^{\text{tree}}$
and $\zeta_{r}^{\text{tree}}$ do not couple
to the first Higgs doublet.

The equation for ${\vec V}_s$,
\begin{equation}
  {\vec V}_s^{\top}M_D^{\top}M_R^{-1}M_D {\vec V}_s=m_s^{\text{tree}}\, ,
\end{equation}
gets solved taking
\begin{equation}
  M_D = m_D {\vec V}_s^{\dagger}\,,
\end{equation}
where $m_D$ is the ``length'' of $M_{D}$
\begin{equation}
\label{V-conditions-H1-mD}
m_D^2 \, := \, M_{D}\cdot M_{D}^{\dagger}\, = \, M_R m_s^{\text{tree}}
\enspace ,
\end{equation}
and corresponds to the Dirac mass term of the effective $2\times 2$ seesaw 
between $\zeta_{s}^{\text{tree}}$ and $\zeta_{4}^{\text{tree}}$.
Using the notation of $M_D$, eq.~(\ref{M_D}), we can express the
Yukawa coupling $\Delta_1$ as
\begin{equation}
\label{Delta1}
  \Delta_{1}
=
  \frac{\sqrt{2}}{v} M_{D} 
=
  \frac{\sqrt{2}}{v} m_{D} \vec{V}_{s}^{\dagger} 
\enspace .
\end{equation}

We would like to write $\Delta_2$ in terms of the
vectors $\vec{V}_{i}$ as well.
We can assume\footnote{Since there are two Yukawa couplings that couple 
the fermionic singlet $\nu_{R}$ to the three generations of neutral leptons, 
they can be viewed as two 3-vectors in generation space. But two 3-vectors
always have a single 3-vector that is orthogonal to both of them. 
This orthogonal state corresponds to the massless neutrino state 
$\zeta_{o}=\zeta_{o}^{\text{tree}}$
and is therefore the justification of our assumption.
It is our choice to consider specific intermediate neutrino states:
(1)~the state $\zeta_{s}^{\text{tree}}$ is aligned
to one Yukawa coupling, eq.~(\ref{Delta1}), and
(2)~one state, $\zeta_{o}$, is orthogonal to the other two states.
The former is affected by the mixing
due to $R_{3}$, eq.~(\ref{R-def}), and the later is not.
}
that the massless
neutrino state $\zeta_{o}$ does not couple to the second Higgs
doublet, either:
\begin{equation}
\label{V-conditions-H2}
  \Delta_{2} \cdot \vec{V}_{o}^{} 
= 
  0
\enspace .
\end{equation}
This condition ensures that the lightest neutrino only couples to the
electroweak sector.
Then we can express $\Delta_2$ in terms of the parameters $d$ and 
$d^{\prime}$ and the vectors $\vec{V}_{i}$ as
\begin{equation}
\label{Delta2}
  \Delta_{2} 
=: 
  d \vec{V}_{r}^{\dagger} 
+ d^{\prime} \vec{V}_{s}^{\dagger} 
\enspace ,
\end{equation}
where we choose the phase of $\vec{V}_{r}$ in such a way, that the
coefficient $d$ becomes real and positive. The coefficient
$d^{\prime}$ may be a complex number. Our goal is to express these
coefficients $d$ and $d^{\prime}$ in terms of the other model parameters.

The neutrino mass matrix, corrected for 1-loop contributions written in
eq.~(\ref{corrections}), gives an effective $3 \times 3$-matrix
\begin{eqnarray}
\mathcal{M}_{\nu}
 &=&
\mathcal{M}_{\nu}^{\text{tree}} + \delta M_L
\enspace ,
\label{M3x3-one-loop}
\end{eqnarray}
which has to be diagonalized like eq.~(\ref{Mtotal}). This diagonalization
gives a vanishing neutrino mass $m_{o} = 0$ and two positive
masses $m_{s}$ and $m_{r}$, which can provide the two measured
neutrino mass squared differences.
Note, that $m_{s}$ can differ from the tree-level value
$m_{s}^{\text{tree}}$ obtained from the diagonalization of
eq.~(\ref{M3x3-tree}).

The tree-level diagonalization matrix $V$ partially diagonalizes
the effective light neutrino mass matrix $\mathcal{M}_{\nu}$, 
eq.~(\ref{M3x3-one-loop}),
and we see explicitly, that it is rank~2:
\begin{equation}
\label{Mnu-eff}
  V^{\top} \mathcal{M}_{\nu} \, V
= 
  \left( \begin{array}{ccc}
  0 & 0 & 0 \\ 0 & a & b \\ 0 & b & c
  \end{array} \right)
=: 
  \left( \begin{array}{ccc}
  0 & 0 & 0
 \\ \begin{array}{c} 0 \\ 0 \end{array} 
  & \multicolumn{2}{c}{\mathcal{M}_{2\times 2}}
  \end{array} \right)
\enspace ,
\end{equation}
with 
\begin{eqnarray}
\label{Mnu-eff-a}
  a 
&=& 
 d^{2} f_{1}
\enspace ,
\\
\label{Mnu-eff-b}
  b 
&=& 
  d^{\prime} d f_{1}
+ d \sfrac{\sqrt{2} m_{D}}{v} f_{2}
\enspace ,
\\
\label{Mnu-eff-c}
  c 
&=& 
  d^{\prime 2} f_{1}
+ 2 d^{\prime} \sfrac{\sqrt{2} m_{D}}{v} f_{2}
+ \sfrac{2 m_{D}^{2}}{v^{2}} f_{3}
\enspace ,
\end{eqnarray}
where
\begin{eqnarray}
\label{f1}
  f_{1}
&=& 
  \sum_{k=1}^{3}
  [ ( b_{k} )_{2} ]^{2} L(m_{H^0_k}^{2})
\enspace ,
\\
\label{f2}
  f_{2}
&=& 
  \sum_{k=1}^{3}
  [( b_{k} )_{2} ( b_{k} )_{1} ] L(m_{H^0_k}^{2})
\enspace ,
\\
\label{f3}
  \tilde{f}_{3} 
&=& 
  3 L(m_{Z}^{2})
+ \sum_{k=1}^{3}
  [ ( b_{k} )_{1} ]^{2} L(m_{H^0_k}^{2})
\enspace ,
\end{eqnarray}
and
\begin{eqnarray}
\label{def-L(m)}
  L(m^{2}) 
:= 
  \frac{1}{32\pi^{2}} 
  \frac{m^{2}}{M_{R}}
  \ln\left[\frac{M_{R}^2}{m^2}\right]
\enspace .
\end{eqnarray}
$f_{3}$ is defined to contain the tree-level contribution, too:
\begin{eqnarray}
\label{def-f3-from-tilde}
  f_{3} := \tilde{f}_{3} - \sfrac{v^{2}}{2 M_{R}}
\enspace .
\end{eqnarray}
The values of $a$, $b$, $c$, $\tilde{f}_3$, and $f_i$ ($i=1,2,3$) are
complex in the general case, as can be seen from the complex entries
in the vectors $b_{k}$, eq.~(\ref{bVectSetTh}). If the Higgs potential 
is CP-conserving, the entries in the vectors $b_{k}$,
table~\ref{table1}, become either real 
or purely imaginary, hence giving real functions $f_1$ and $f_3$.

For getting the masses and the mass eigenstates, we use the 
Takagi Factorization~\cite{Hahn:2006hr} for 
eq.~(\ref{Mnu-eff}) with the unitary matrix $R_{3}$ 
\begin{equation}
\label{Mnu-eff-diag}
  R_{3}^{\top} V^{\top} \mathcal{M}_{\nu} \, V \, R_{3}
= 
  \text{diag} \left( 0 
  , \text{diag} \left( R_{2}^{\top} \mathcal{M}_{2\times 2} \, R_{2} \right)
  \right)
= 
  \text{diag} \left( 0 , m_{r} , m_{s} \right)
\enspace .
\end{equation}
$R_{3}$ only mixes the massive states $\zeta_{r}$ and $\zeta_{s}$, hence 
we can parameterize it as
\begin{equation}
\label{R-def}
  R_{3}
= 
  \left( \begin{array}{cc}
  e^{i\alpha_{o}} & 0 \\ 0 & R_{2}
  \end{array} \right)
\quad\text{and}\quad
  R_{2}
= 
  \left( \begin{array}{cc}
    \cos\beta
   & - e^{i\gamma} \sin\beta 
 \\ e^{-i\gamma} \sin\beta 
   & \cos\beta
  \end{array} \right)
\cdot 
  \left( \begin{array}{cc}
   e^{i\alpha_{r}} & 0
 \\ 0 & e^{i\alpha_{s}}
  \end{array} \right)
\enspace ,
\end{equation}
where the parameters $\beta$ and $\gamma$ describe effectively 
only a $2\times 2$ unitary matrix. The phases $\alpha_{i}$ have
to be determined together with the possible Majorana phases of the 
light neutrinos. 
$\beta$ and $\gamma$ can be determined from the linear relation
$R_{2}^{\top} \mathcal{M}_{2\times 2} = 
  \text{diag} \left( m_{r} , m_{s} \right) R_{2}^{\dagger}$
with the abbreviations
\begin{eqnarray}
\label{pq} 
  p = \sfrac{1}{2} ( a^{*} a - c^{*} c )
\enspace\text{, }\quad
  \tilde{q} = a^{*} b + b^{*} c 
\enspace\text{, and}\quad
  q = | \tilde{q} | = | a^{*} b + b^{*} c |
\end{eqnarray}
to be 
\begin{eqnarray}
\label{tb} 
  \tan\beta
=
  t_{\beta}
=
  \frac{ q }{ p \pm \sqrt{p^{2} + q^{2}} }
=
  \frac{- p \pm \sqrt{p^{2} + q^{2}} }{ q }
\enspace ,
\end{eqnarray}
and
\begin{eqnarray}
\label{eig} 
  e^{i\gamma}
&=& 
  \frac{\tilde{q}}{q}
\, = \,
  \frac{a^{*} b + b^{*} c}{| a^{*} b + b^{*} c |} 
\enspace .
\end{eqnarray}
The masses are most easily obtained as the eigenvalues of the squared
matrix
\begin{equation}
\label{M2x2square}
  A 
=
  \mathcal{M}_{2\times 2}^{\dagger} \mathcal{M}_{2\times 2}^{}
= 
  \left( \begin{array}{cc}
  a^{*} a + b^{*} b & a^{*} b + b^{*} c
 \\ a b^{*} + b c^{*} & b^{*} b + c^{*} c 
  \end{array} \right)
= 
  \left( \begin{array}{cc}
  s + p & \tilde{q}
 \\ \tilde{q}^{*} & s - p
  \end{array} \right)
\enspace ,
\end{equation}
where $s=\sfrac{1}{2} \text{Tr}[A] = \sfrac{1}{2} ( m_{r}^{2} + m_{s}^{2} )$.
The masses then are given by
\begin{eqnarray}
\label{mr2ms2} 
  m^{2}_{r,s} 
&=& 
  s \mp \sqrt{s^2 - \text{det}[A]}
\, = \,
  s \mp \sqrt{s^2 - [ s^{2} - p^{2} - \tilde{q}^{*} \tilde{q} ]}
\, = \,
  s \mp \sqrt{p^2 + q^{2}}
\enspace .
\end{eqnarray}
The phases $\alpha_{r}$ and $\alpha_{s}$ have to be extracted from 
the relation linear in $\mathcal{M}_{2\times 2}$, eq.~(\ref{Mnu-eff-diag}), 
\begin{eqnarray}
\label{mrt} 
  e^{-2i\alpha_{r}} m_{r} 
&=& 
  \frac{a + 2 b \, t_{\beta} e^{-i\gamma} + c \, t_{\beta}^2 e^{-2i\gamma} }
       {1 + t_{\beta}^2}
\enspace ,
\end{eqnarray}
and
\begin{eqnarray}
\label{mst} 
  e^{-2i\alpha_{s}} m_{s} 
&=& 
  \frac{a \, t_{\beta}^2 e^{2i\gamma} - 2 b \, t_{\beta} e^{i\gamma} + c }
       {1 + t_{\beta}^2}
\enspace ,
\end{eqnarray}
as they drop out in the squared relations.
One additional relation for the phases can be obtained from the determinant
\begin{equation}
\label{det-M2x2}
  a c - b^{2} 
= 
  \det \mathcal{M}_{2\times 2} 
= 
  \det [ R_{2}^{*} \text{diag}( m_{r} , m_{s} ) R_{2}^{\dagger} ]
=
  e^{-2i\alpha_{r}} m_{r} \, e^{-2i\alpha_{s}} m_{s}
\enspace , 
\end{equation}
which can serve as a numerical consistency condition for the 
extraction of the phases from eqs.~(\ref{mrt}) and (\ref{mst}).

With the rotation matrix $R_{3}$, eq.~(\ref{R-def}), we have now the 
transformation matrix between the flavour eigenstates $\nu_{L}$ and 
the light neutrino mass eigenstates $\zeta$
\begin{eqnarray}
\label{nuL-to-masseigenstates} 
  \nu_{L} 
=
  V \, R_{3} \, \zeta
=
  V_{\mathrm{PMNS}} \, \zeta
\enspace ,
\end{eqnarray}
which allows us to identify our vectors $\vec{V}_{i}$ with 
columns of the PMNS matrix, eq.~(\ref{PMNS}). Since we chose to
identify $\zeta_{o}$ with the massless neutrino, $\zeta_{r}$
with the neutrino, that gets its mass only with radiative corrections, 
and $\zeta_{s}$ with the neutrino that already has a mass from the 
seesaw mechanism, we have to take the corresponding columns from 
the PMNS matrix to determine our vectors, that we want to use for the 
definition of the Yukawa couplings: 
\begin{eqnarray}
\label{Vo-from-PMNS} 
  \vec{V}_{o} 
&=& 
  ( V_{\mathrm{PMNS}} )_{o} e^{-i\alpha_{o}}
\enspace ,
\\
\label{Vr-from-PMNS} 
  \vec{V}_{r} 
&=& 
  \cos\beta \, ( V_{\mathrm{PMNS}} )_{r} e^{-i\alpha_{r}} 
- e^{-i\gamma} \sin\beta  \, ( V_{\mathrm{PMNS}} )_{s} e^{-i\alpha_{s}}  
\enspace ,
\\
\label{Vs-from-PMNS} 
  \vec{V}_{s} 
&=& 
  e^{i\gamma} \sin\beta \, ( V_{\mathrm{PMNS}} )_{r} e^{-i\alpha_{r}} 
+ \cos\beta  \, ( V_{\mathrm{PMNS}} )_{s} e^{-i\alpha_{s}} 
\enspace ,
\end{eqnarray}
where the numbers for the columns have to be taken according to 
table~\ref{index-table}.

Relating the measured mass squared differences $\Delta m^2_{21}$
and $\left|\Delta m_{31}^{2}\right|$ to the masses of the three light 
neutrinos $m_{i}$ we can express the parameters of the Yukawa couplings
of the neutrinos to the second Higgs doublet by measured quantities. 

Inserting the definitions of the matrix elements $a$, $b$, and $c$
(eqs.~(\ref{Mnu-eff-a})-(\ref{Mnu-eff-c})) into the relation
eq.~(\ref{det-M2x2}), we can derive:
\begin{eqnarray}
\label{getting-d2} 
  e^{-2i(\alpha_{r}+\alpha_{s})} m_{r} m_{s}
&=& 
  a c - b^{2}
= 
  d^{2} f_{1}
  ( d^{\prime 2} f_{1}
  + 2 d^{\prime} \sfrac{\sqrt{2} m_{D}}{v} f_{2}
  + \sfrac{2 m_{D}^{2}}{v^{2}} f_{3} 
  )
- ( d^{\prime} d f_{1}
  + d \sfrac{\sqrt{2} m_{D}}{v} f_{2} )^{2}
\nonumber\\
&=& 
  d^{2} f_{1}
  \sfrac{2 m_{D}^{2}}{v^{2}} f_{3} 
- d^{2} \sfrac{2 m_{D}^{2}}{v^{2}} f_{2}^{2}
\nonumber \\
&=&
  d^{2} \,
  \sfrac{2 m_{D}^{2}}{v^{2}} [ f_{1} f_{3} - f_{2}^{2} ]
\enspace .
\end{eqnarray}
Taking the modulus we get the functional expression for $d^2$
\begin{eqnarray}
\label{d2dep} 
  d^{2}
& = &
  d^{2} [ v^{2} ; m_{H^{0}_{i}}, \text{s}_{\vartheta} 
        ; m_{r} , m_{s} , m_{4} 
        ; m_{D}^{2} ]
\, = \, 
\frac{v^{2}}{2 m_{D}^{2}}
\frac{m_{r} m_{s}}{| f_{1} f_{3} - f_{2}^{2} |}
\enspace ,
\end{eqnarray}
where we treat $m_{D}^{2}$ as a free parameter, since in general
$m_{s} \neq m_{s}^{\text{tree}}$. 

To get an expression for the modulus of $d^{\prime}$
\begin{eqnarray}
\label{dpdef} 
  d^{\prime}
& = &
  |d^{\prime}| 
  e^{i \phi^{\prime} }
\enspace ,
\end{eqnarray}
we take the trace of $[\text{eq.~}(\ref{Mnu-eff-diag})]\cdot
[\text{eq.~}(\ref{Mnu-eff-diag})]^{\dagger}$, which gives
$m_{r}^{2} + m_{s}^{2}$ on the r.h.s.\ and
$( |a|^{2} + |b|^{2} ) + ( |b|^{2} + |c|^{2} )$ on the
l.h.s. By reversing the sides, we write
a fourth order polynomial in $|d^{\prime}|$:
\begin{eqnarray}
\label{getting-dp}
m_{r}^{2} + m_{s}^{2}
& = &
  d^{4} |f_{1}|^{2}
+ 2 d^{2} | d^{\prime} f_{1} + \sfrac{\sqrt{2} m_{D}}{v} f_{2} |^{2}
+ | d^{\prime 2} f_{1}
  + 2 d^{\prime} \sfrac{\sqrt{2} m_{D}}{v} f_{2}
  + \sfrac{2 m_{D}^{2}}{v^{2}} f_{3} 
  |^{2}
\nonumber\\
&=& 
  a_{4} |d^{\prime}|^{4}
+ a_{3} |d^{\prime}|^{3}
+ a_{2} |d^{\prime}|^{2}
+ a_{1} |d^{\prime}|
+ \tilde{a}_{0} 
\enspace .
\end{eqnarray}
The general expressions for the coefficients $a_{i}$ are simpler in our
CP conserving case with the $b$-vectors having the form of
eq.~(\ref{bVectSetTh}). In this case, the values of $f_{1}$ and
$f_{3}$, given in eqs.~(\ref{f1}) and (\ref{def-f3-from-tilde}), are
real numbers, leading to
\begin{eqnarray}
\label{a4} 
  a_{4} 
& = &
  f_{1}^{2}
\\
\label{a3} 
  a_{3} 
& = &
  4 \sfrac{\sqrt{2} m_{D}}{v} f_{1} 
  \left[ \text{Re}[ f_{2} ] \cos\phi^{\prime} 
       + \text{Im}[ f_{2} ] \sin\phi^{\prime} 
  \right]
\\
\label{a2} 
  a_{2} 
& = &
  2 d^{2} f_{1}^{2}
+ 4 \sfrac{2 m_{D}^{2}}{v^{2}} |f_{2}|^{2}
+ 2 \sfrac{2 m_{D}^{2}}{v^{2}} f_{1} f_{3} ( 2 \cos^{2}\phi^{\prime} - 1 )
\\
\label{a1} 
  a_{1} 
& = &
  4 \sfrac{\sqrt{2} m_{D}}{v} 
  \left(
    [ d^{2} f_{1} + \sfrac{2 m_{D}^{2}}{v^{2}} f_{3} ]
       \text{Re}[ f_{2} ] \cos\phi^{\prime} 
+
    [ d^{2} f_{1} - \sfrac{2 m_{D}^{2}}{v^{2}} f_{3} ]
    \text{Im}[ f_{2} ] \sin\phi^{\prime} 
  \right)
\\
\label{a0} 
  a_{0}
&=&
  \tilde{a}_{0} 
- [ m_{r}^{2} + m_{s}^{2} ] 
\, = \, 
  d^{4} f_{1}^{2}
+ 2 d^{2} \sfrac{2 m_{D}^{2}}{v^{2}} |f_{2}|^{2}
+ \sfrac{4 m_{D}^{4}}{v^{4}} f_{3}^{2}
- [ m_{r}^{2} + m_{s}^{2} ] 
\enspace .
\end{eqnarray}
The value of
$|d^{\prime}|$ is then given as a real positive solution to the
fourth order equation
\begin{eqnarray}
\label{4o-eq-dp} 
  a_{4} |d^{\prime}|^{4}
+ a_{3} |d^{\prime}|^{3}
+ a_{2} |d^{\prime}|^{2}
+ a_{1} |d^{\prime}|
+ a_{0} 
&=&
  0
\enspace .
\end{eqnarray}
Therefore $|d^{\prime}|$ has the dependence
\begin{eqnarray}
\label{dpdep} 
  |d^{\prime}|
& = &
  |d^{\prime}| 
  [ v^{2} ; m_{H^{0}_{i}}, \text{s}_{\vartheta} 
  ; m_{r} , m_{s} , m_{4} ; m_{D}^{2} 
  ; \phi^{\prime} ]
\enspace .
\end{eqnarray}
In order to find a real and positive solution, the value of the phase 
$\phi^{\prime}=\mathrm{arg}(d^{\prime})$ can be restricted.

Using $d$ and $|d^{\prime}|$ we have analytically replaced two
of our input parameters with the neutrino masses. The replacement 
of the Yukawa couplings to the fermionic singlet is done by 
eqs.~(\ref{Delta1}) and (\ref{Delta2}), using the determined parameters
$d$ and $|d^{\prime}|$, and the vectors $\vec{V}_{i}$,
eqs.~(\ref{Vo-from-PMNS}), (\ref{Vr-from-PMNS}), and 
(\ref{Vs-from-PMNS}). 

\subsection{The choice of the initial parameters for the numerical analysis}

The primary choice are the parameters appearing in the Lagrangian, which 
define the model. More convenient is a choice, where some of the parameters 
can be directly related to measured quantities. At the tree-level we 
achieve this simplification, by using part of the guidelines 
in~\cite{Ofreid:2hdmwork2018,Ogreid:2018bjq}. Taking for the 
2HDM potential the masses of the Higgs particles and their mixing angles
and ignoring the 2HDM parameters that do not enter our calculations  
we get the list:
\begin{eqnarray}\label{parameterListLagrangian-tree}
\{ 
  p_{i,\text{SM}} ; m^{2}_{H_{k}^{0}} , \text{s}_{12} , \text{s}_{13} 
; M_{R} , \Delta_{j} , \Gamma_{2}
\}
\enspace .
\end{eqnarray}
This is the general and basic parameter list for scans of the parameter
space of the model. 

When using our analytical result for the neutrino masses, we can reduce 
this parameter list by replacing the Yukawa couplings $\Delta_{j}$ by
their values, eqs.~(\ref{Delta1}) and (\ref{Delta2}). 
This means, we have to take the neutrino masses as input, also 
replacing $M_{R}$ with $m_{4}$, as the seesaw mechanism, 
eq.~(\ref{Mtotal}), gives the relation 
\begin{eqnarray}\label{seesaw-MR}
  M_{R} = m_{4} - m_{s} \approx m_{4}
\enspace ,
\end{eqnarray}
even if we do not expect to measure the mass of the heavy neutrino.
Since for simplicity we assumed a CP-conserving Higgs potential, we can 
also simplify the mixing angles of the neutral Higgs bosons, 
either $\text{s}_{12}$ or $\text{s}_{13}$, as given
in table~\ref{table1}, to a
single angle $\text{s}_{\beta-\alpha}$. 
That leaves us with the same parameter list as the dependencies of
$|d^{\prime}|$, eq.~(\ref{dpdep}). From these the only parameter, that 
does not have an immediate physical meaning is $m_{D}^{2}$.
We know from the tree-level seesaw relation eq.~(\ref{V-conditions-H1-mD})
that
\begin{eqnarray}\label{seesaw-MD}
  m_{D}^{2} / M_{R} = m_{s}^{\text{tree}}
\enspace ,
\end{eqnarray}
but that does not tell us the value of $m_{s}^{\text{tree}}$. 
Assuming that our model has a sensible loop expansion, we can make
the educated guess, that $m_{s}^{\text{tree}}$ should be of the same
order as the physical mass $m_{s}^{}$, which we identify with one 
of the light neutrino masses. For simplicity we parameterize the 
change from $m_{s}^{\text{tree}}$ to $m_{s}^{}$ as a multiplicative 
parameter 
\begin{equation}
\label{def_scaling}
  m_{D}
=
 \sqrt{m_{s}^{\text{tree}}  M_{R}}
:=
  \lambda_{D}
  \sqrt{m_{s}^{} m_{4}^{}}
\enspace , 
\end{equation}
that we call $\lambda_{D}$, as it enters at the place of $m_{D}^{2}$.

Since in our model the lightest neutrino stays massless,
the measured neutrino squared mass differences~\cite{deSalas:2017kay},
\begin{equation}
 \Delta m_{21}^{2} = m_{2}^{2} - m_{1}^{2}
 \enspace ,
\quad\text{and}\quad
 \left|\Delta m_{31}^{2}\right| = \left|m_{3}^{2} - m_{1}^{2}\right|
\enspace ,
\end{equation}
give the estimates of the light neutrino masses for the normal hierarchy
\begin{equation} \label{NHmasses}
 m_{o} = m_{1} = 0, 
\quad
 m_{2} = \sqrt{\Delta m_{21}^{2}}
,
\quad\text{and}\quad
 m_{3} = \sqrt{\left|\Delta m_{31}^{2}\right|}
\enspace ,
\end{equation}
and for the inverted hierarchy
\begin{equation} \label{IHmasses}
 m_{1} = \sqrt{\left|\Delta m_{31}^{2}\right|}, 
\quad
 m_{2} = \sqrt{\Delta m_{21}^{2} + \left|\Delta m_{31}^{2}\right|}
, \quad\text{and}\quad
 m_{o} = m_{3} = 0
\enspace.
\end{equation}
Using the assignments of Table~\ref{index-table} we connect the 
value of the parameter $m_{D}^{}$ to the masses obtained in
eqs.~(\ref{NHmasses}) and (\ref{IHmasses}): 
\begin{eqnarray}
  m_{D}^{}
&=&
  \lambda_{D} \sqrt{m_4^{} m_{3}^{}}
= 
  \lambda_{D} \sqrt{m_4 \sqrt{\left|\Delta m_{31}^{2}\right|}}
\quad\text{for NH}, \\
  m_{D}^{}
&=&
  \lambda_{D} \sqrt{m_4^{} m_{2}^{}}
= 
  \lambda_{D} \sqrt{m_4 \sqrt{\Delta m_{21}^{2}}}
\quad\text{for }\overline{\text{NH}}, \\
  m_{D}^{}
&=&
  \lambda_{D} \sqrt{m_4^{} m_{2}^{}}
= 
  \lambda_{D} 
  \sqrt{m_4 \sqrt{\Delta m_{21}^{2} + \left|\Delta m_{31}^{2}\right|}}
\quad\text{for IH}, \\
  m_{D}^{}
&=&
  \lambda_{D} \sqrt{m_4^{} m_{1}^{}}
= 
  \lambda_{D} \sqrt{m_4 \sqrt{\left|\Delta m_{31}^{2}\right|}}
\quad\text{for }\overline{\text{IH}}.
\end{eqnarray}
We assume that the one loop corrections do not invalidate the tree level
assumptions for the seesaw. This allows us to restrict the scaling parameter
to the range $\frac{1}{2} \le \lambda_{D} \le 2$.

Having made these adjustments to the parameters, we arrive at three separate
sets of parameters for our analysis. (1)~There are several input parameters
that (a)~are not affected by our calculations, like the Standard Model
parameters $p_{i,\text{SM}}$, (b)~the parameters of the 2HDM, that do
not enter in the calculation of the neutrino masses, like the Higgs potential
parameters $\lambda_{i}$
that do not enter the tree-level Higgs masses, and (c)~the
Yukawa coupling of the second Higgs doublet to the charged fermions. 
We summarize the first set of parameters with the name
\begin{eqnarray}\label{parList-SM}
\tilde{p}_{i,\text{SM}} = 
\{ 
  p_{i,\text{SM}} , \text{ some }\lambda_{i}\text{'s } , \Gamma_{2} 
\}
\enspace .
\end{eqnarray}
Then (2)~there are the parameters that are always used as input for
the calculation of the neutrino masses, 
\begin{eqnarray}\label{parList-input}
\{ 
  m_{H^{0}_{i}} , \text{s}_{\vartheta} 
, m_{4} , \lambda_{D} , \phi^{\prime} 
\}
\enspace ,
\end{eqnarray}
where we use the same symbol $\text{s}_{\vartheta}$ for both angles 
$\text{s}_{\vartheta_{1j}}$, table~\ref{table1},
as we have only one non-vanishing
mixing angle due to our simplification of 
taking only a CP-conserving Higgs sector.
Comparing to~\cite{Haber:2006ue} we have
$\text{c}_{\vartheta} = \text{s}_{\beta-\alpha}$.

It is easy to generalize our calculation to a CP non-conserving
Higgs potential. We do not expect additional difficulties.
In principle, just the intermediate parameter $f_{1}$, defined
in eq.~(\ref{f1}), will become complex.
The  biggest difficulty would be to present our results in the 
extended parameter space,
while the conclusions of our study would not change.

And (3)~there are parameters, that can be both input and output
of our calculations. For example, the neutrino parameters
\begin{eqnarray}\label{parList-neutrino}
\{ 
  \Delta m_{21}^{2} , |\Delta m_{31}^{2}| , V_{\text{PMNS}}
\}
\enspace ,
\end{eqnarray}
are an input, if we use the procedure of this section. But they
become an output, if we stay with the Lagrangian parameters 
$\{ \Delta_{j} \}$ as input, as is the starting point of
Section~\ref{framework}.

\section{Numerical analysis}
\label{NumRes}

Usually, the model parameters are the quantities
that are defined in the Lagrangian, and the predictions are the
measureable quantities,
like masses and cross sections. Therefore,
the Yukawa couplings and the parameters of the Higgs potential
should be treated as our input parameters. Since our interest in
the Higgs sector is limited, we take the masses and the mixing angle
of the neutral Higgs bosons as input parameters.

Considerations are different in the neutrino sector.
On one hand, using the approximations of
Grimus and Lavoura~\cite{Grimus:2002nk}
we treat the Yukawa couplings $\Delta_{j}$, eq.~(\ref{Yukawa}),
together with the Majorana mass $M_{R}$,
the Higgs masses and the Higgs mixing angle
as input parameters and ``predict'' the neutrino masses and mixings.
Analysing the model in this way, one can fit the input parameters to obtain 
the physically measured neutrino mass differences and the neutrino mixing 
matrix. For this approach
one has to construct a minimization function that allows 
to find the global minimum, which should give the model parameters that 
correspond to the physically measured values. 

On the other hand we can use our analytic results for the neutrino masses 
to directly determine the Yukawa couplings $\Delta_{j}$,
eqs.~(\ref{Delta1}) and (\ref{Delta2}), from the measured neutrino
parameters and other input parameters via evaluation of the
orthonormal vectors $\vec{V}_{i}$. We determine $d$, $d^{\prime}$,
and $R_{3}$ from eqs.~(\ref{d2dep}), (\ref{dpdep}), and (\ref{R-def}),
and relate $\vec{V}_{i}$ to the measured neutrino mixing matrix
by eqs.~(\ref{Vo-from-PMNS}), (\ref{Vr-from-PMNS}), and (\ref{Vs-from-PMNS}).
Note that the numerical calculations use the
best fit values~\cite{deSalas:2017kay} of the oscillation angles 
$\theta_{12}$, $\theta_{13}$, $\theta_{23}$,
and the Dirac phase $\delta_{CP}$ as input for the PMNS matrix.
Using thus obtained values of the Yukawa couplings $\Delta_{j}$,
we can again go back and 
``predict'' the neutrino masses and mixings, compare the result
with the measured masses, and hopefully save a lot of time by having
to sample over a much smaller parameter space:
we have to vary only one phase $\phi^{\prime}$ and the scaling parameter 
$\lambda_D$, eq.~(\ref{def_scaling}), instead of $6$ complex
entries in $\Delta_{j}$, eq.~(\ref{Yukawa}).
This is the procedure we adopt in first subsection,
\ref{PartCaseSect}, to check the consistency of our approach. 

The second subsection discusses the allowed parameter space by showing 
various distributions of parameters and interpreting the restrictions 
that can be seen in the plots.
In the third subsection we argue that our analytical approach 
has advantages over the ``blind'' systematic scanning of the allowed
parameters that go beyond the simple saving of computer time. 

All the numerical analysis was performed using data points of
the Higgs sector that were subjected to additional theoretical and 
experimental constraints similar to~\cite{AntonData,AK}, as described in 
appendix~\ref{Appendix-2HDM}: 
the CP-conserving 2HDM potential should be stable, 
guarantee tree-level unitarity of the $S$ matrix, 
be bounded from below, have a global minimum, 
and fulfill the experimental restrictions of 
the Peskin-Takeuchi $S$, $T$, and $U$ parameters. 
Additionally, the SM Higgs boson has the mass 
$m_{h}=125.18$~GeV~\cite{PDG2018} and the masses $m_{H}$ and $m_{A}$ 
of the other two neutral Higgs bosons vary in the range from 
$m_{h}$ to $3000$~GeV. The mixing angle between $h^{0}$ and $H^{0}$
varies in the range from $-\pi/2$ to $\pi/2$, 
where we assume that $h^{0}$ to corresponds to the SM Higgs boson.
%

%
%
Progress in the experimental particle physics program at the LHC
limits the Higgs sector parameter space.  Haller
\textit{et al}~\cite{Haller:2018nnx} summarized the restrictions
on the 2HDM parameters.
We checked that most of the Higgs potential points that
pass our theoretical restrictions will also fulfill the more
restrictive and specialized constraints of the ``typed'' 2HDM
models.

The behavior of some numerical solutions in our model is illustrated
using the benchmark point B1 of~\cite{Hespel:2014sla}.  Originally,
this point was used for the 2HDM type-II studies~\cite{Baglio:2014nea} 
(there it was named H-1) and was recently
excluded~\cite{Haller:2018nnx}. However, the values would still be
valid for the 2HDM type-I model~\cite{Haller:2018nnx,Arbey:2017gmh}.
Since we make no distinction for the type of the 2HDM model, we use
this point having updated the mass of the lightest Higgs boson
$h^{0}$. More details are provided below.

\subsection{Numerical consistency of the model}
\label{PartCaseSect}
As the first test of our approach we calculate the Yukawa couplings, 
eqs.~(\ref{Delta1}) and (\ref{Delta2}), using the input parameters
eq.~(\ref{parList-input}). For that we have to calculate also the 
matrix $R_{3}$, eq.~(\ref{R-def}), in order to use the correct
orthonormal basis $( \vec{V}_{o} , \vec{V}_{r} , \vec{V}_{s} )$ that
defines the Yukawa couplings. These numerical values 
of the Yukawa couplings we treat as input in the sense of 
eq.~(\ref{parameterListLagrangian-tree}) and calculate the masses 
and the mixing matrix between the neutrino mass eigenstates and the interaction
states: as expected, the mass differences agree between input and output. 
For the neutrino mixing angles it makes a difference, whether we adopt our
complicated procedure, described in sec.~\ref{framework:improved couplings}, 
or we just take the measured PMNS matrix
as the orthonormal basis and ignore the difficulty of calculating $R_{3}$.
We get the neutrino mixing angles back in the first case,
whereas in the second case the range of $\phi^{\prime}$, that allows 
solutions to eq.~(\ref{4o-eq-dp}), is reduced to few points where the 
angles of the obtained mixing matrix lie in the $3\sigma$ bands of the 
experimentally allowed values. In the second case it can even happen, 
that we cannot find any values of $\phi^{\prime}$ that allow suitable 
angles of the calculated PMNS matrix.

\subsection{Distributions of the model parameters}
\label{ComplCaseSect}
%
%
\begin{figure}[t]
\begin{center}
\includegraphics[width=0.8\textwidth]{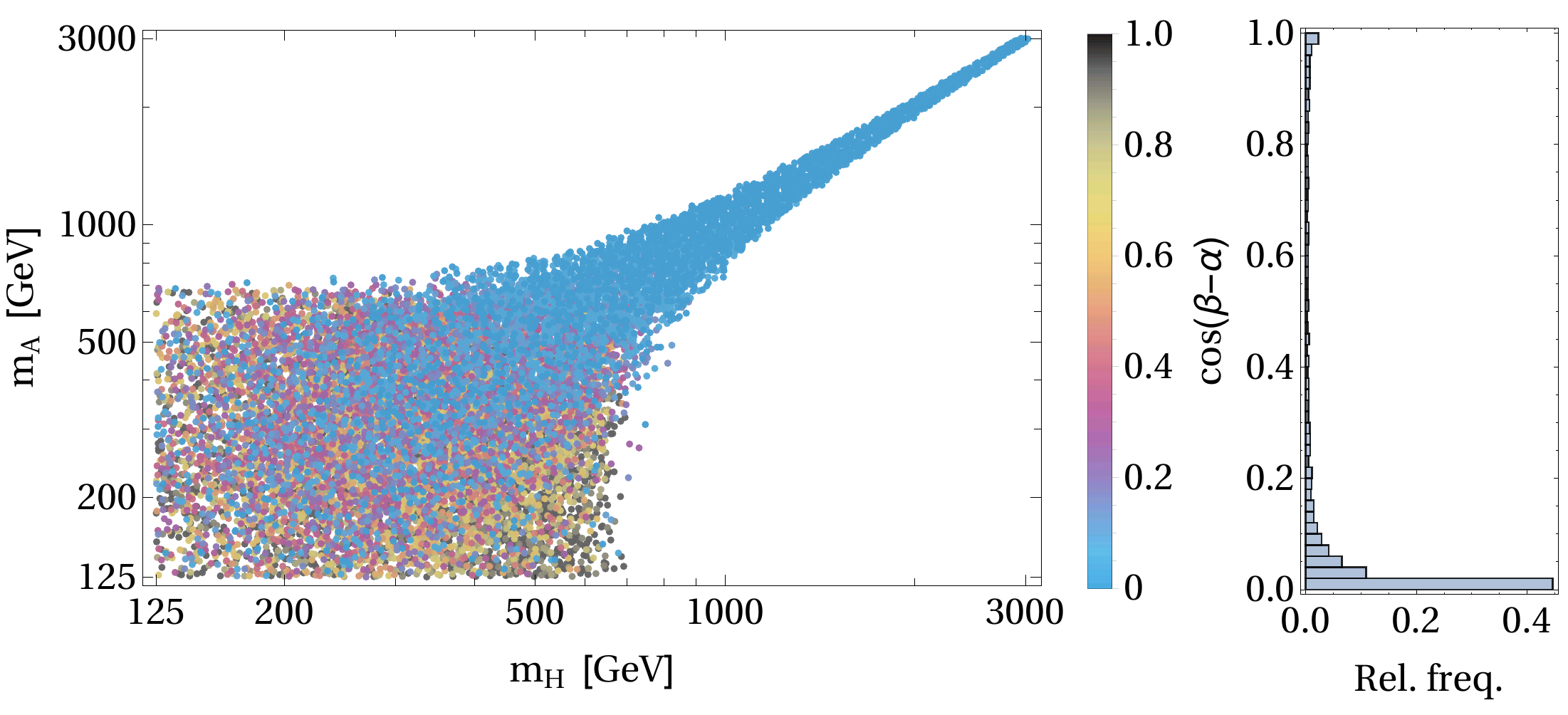}
\vspace{-20pt}
\end{center}
\caption{(Color online) Scatter plot of the tree-level masses of the
heavier Higgs bosons $H^{0}$ and $A^{0}$ together with the color coding
of the cosine of the mixing angle $\beta-\alpha$ between the two CP-even
Higgs bosons $h^{0}$ and $H^{0}$. The relative frequency of the
$\cos(\beta-\alpha)$ values is shown by a histogram on the right.
The plot shows 10.000 points in total. 
}
\label{higgses}
\end{figure}

As a first overview we show the distribution of masses of the heavier 
scalar and pseudoscalar Higgs bosons and
the cosine of the mixing angle $\beta-\alpha$ 
between the two CP-even Higgs bosons $h^{0}$ and $H^{0}$ in 
figure~\ref{higgses}. The allowed Higgs potential points are calculated
with the procedure described in appendix~\ref{Appendix-2HDM}. 
The density of points in $(m_{H},m_{A})$ plane is equalized in
the non-logarithmic scale to have a more uniform representation of
different $m_{H}$ and $m_{A}$ combinations.
Figure~\ref{higgses} clearly shows the restriction 
on the mixing angle $\beta-\alpha$ when the masses become large, 
indicating the onset of the decoupling regime:
$|\beta-\alpha|\rightarrow\pi/2$, when $m_{H},m_{A}\gtrsim700$~GeV.
This agrees with the experimental constraints from the LHC
measurements~\cite{Haller:2018nnx} suggesting
$\cos(\beta-\alpha)\lesssim0.4$.
Nearly all considered points satisfy this limit, as indicated by
the relative frequency distribution of $\cos(\beta-\alpha)$.
%
%
\begin{figure}[t]
\begin{center}
\includegraphics[width=0.6\textwidth]{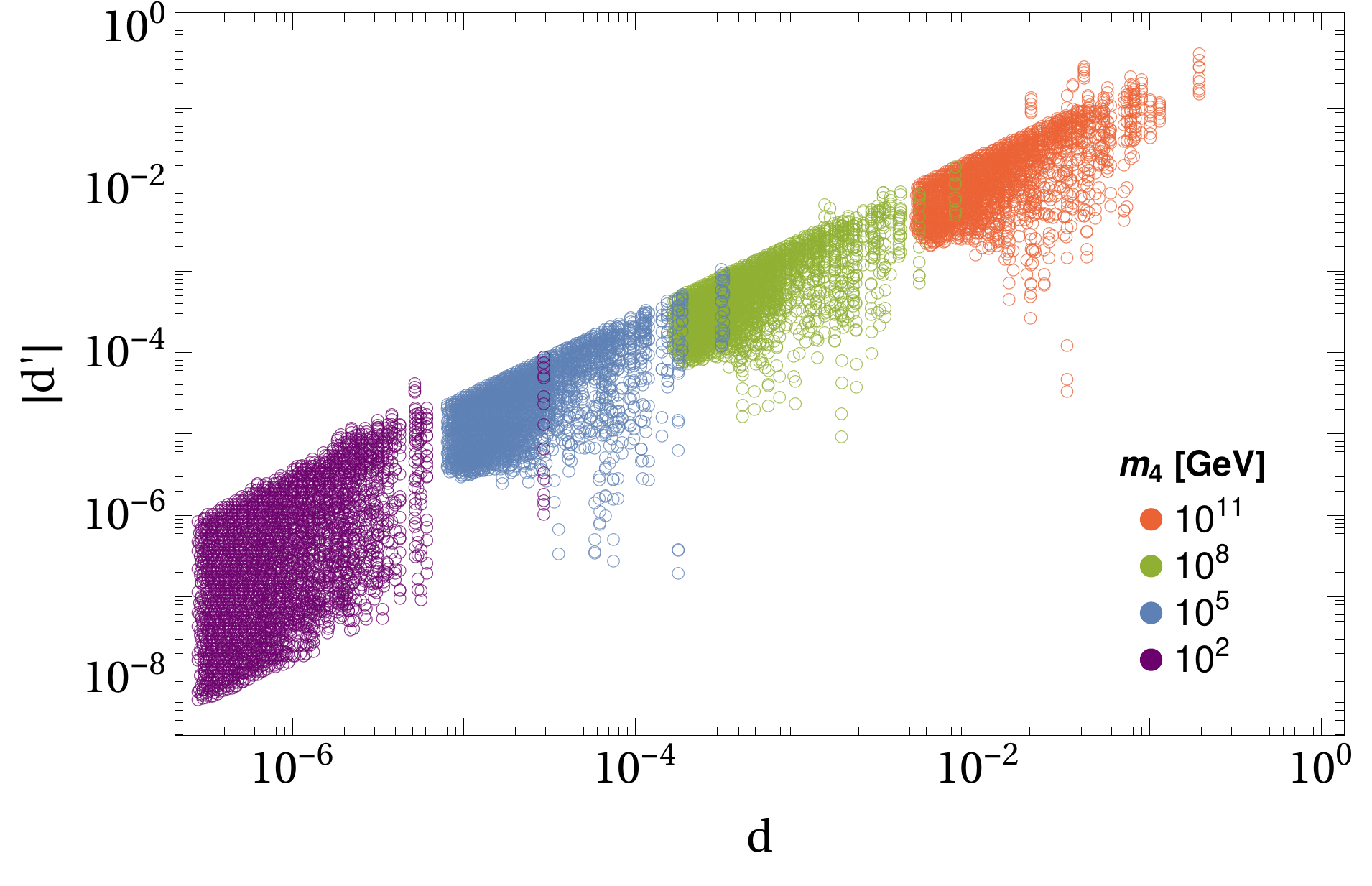}
\vspace{-20pt}
\end{center}
\caption{(Color online) Scatter plot of the values $d$ and $|d^{\prime}|$,
running over 1.000 points in the Higgs sector, figure~\ref{higgses}, 
for specific values of the heavy neutrino mass $m_{4}$. We take
$\lambda_{D} = 1$ and sample  
$\phi^{\prime}$ in the range of allowed values that give a solution to 
the fourth order equation~(\ref{4o-eq-dp}). Each bunch in $m_{4}$ 
has more than 30.000 points. }
\label{dpdm4-plot}
\end{figure}

Figure~\ref{dpdm4-plot} illustrates the spread of values of $d$ and 
$|d^{\prime}|$ coming from the distribution of the Higgs masses and the
mixing angle. The points in
figure~\ref{dpdm4-plot} have $\lambda_{D} = 1$ and are sampled over 
allowed values of $\phi^{\prime}$, but taken only from a reduced set 
of 1.000 Higgs potential parameter points for clarity, as
this reduced set gives a high enough statistical representation. 
These 1.000 points are also evenly distributed in the masses $m_{H}$ 
and $m_{A}$, like figure~\ref{higgses}. The 
sampling over $\phi^{\prime}$ increases the number of points from
1.000 to over 30.000 for each value of $m_{4}$ in figure~\ref{dpdm4-plot}.
The number of points for each value of $m_{4}$ are not exactly equal, 
as few points have less solutions with increasing values of $m_{4}$.
The parameters
$d$ and $|d^{\prime}|$ have an asymptotic scaling $\propto(m_{4})^{4/9}$.

Some features of the ($d$, $|d^{\prime}|$) distribution seen in
figure~\ref{dpdm4-plot} can be understood as follows:
The parameter $d$ has a dependence on Higgs masses expressed
in eqs.~(\ref{d2dep}) and (\ref{f1}) - (\ref{f3}). When masses
$m_H \approx m_A$, the denominator in eq.~(\ref{d2dep}) becomes small,
and the values of $d$ become larger and more scattered.
The sharp ``edges'' in the distribution
(i.e.\ the lower limit for the parameter $d$ for a
fixed value of $m_4$, and the upper limit for the parameter $|d^{\prime}|$
for a fixed $d$ and $m_4$) result from a larger denominator in
eq.~(\ref{d2dep}) and the restrictions in the Higgs sector that lead to
$|m_{H}-m_{A}|\lesssim560$~GeV.
The parameter $|d'|$ has a more complicated dependence on the Higgs masses,
therefore the values are scattered towards both smaller and larger values.
A wider spreading of values occurs in ``exotic'' cases, when three or four
real positive solutions to eq.~(\ref{4o-eq-dp}) exist.
The described features of the parameter $d$ and $|d'|$ distributions
are illustrated below, having discussed the benchmark point and the
Yukawa couplings.

%
%
%
\begin{figure}[t]
\begin{center}
\includegraphics[width=1.0\textwidth]{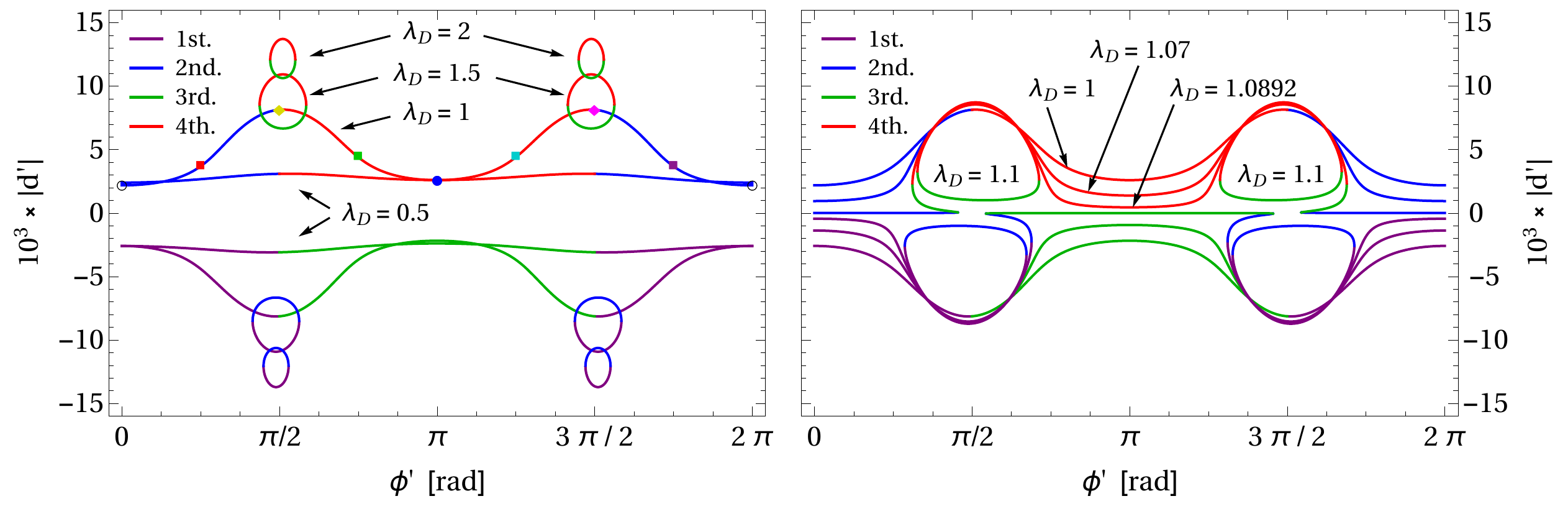}
\vspace{-20pt}
\end{center}
\caption{(Color online) The four solutions for $|d^{\prime}|$ of 
eq.~(\ref{4o-eq-dp}) at different values of
$\phi^{\prime}=\arg(d^{\prime})$ is shown. The dependency changes with
$\lambda_{D}$: a coarse change of $\lambda_{D}$ is shown on the
left, and a finer study of is shown on the right.
$M_{R} \sim m_{4} = 10^{10}\,$GeV is fixed.
The normal hierarchy is assumed.
The parameters of the Higgs sector
are taken from the benchmark point B1~\cite{Hespel:2014sla}: 
$m_{H} = 300\,$GeV, 
$m_{A} = 441\,$GeV, and $\beta-\alpha = 0.522\pi \equiv -0.478\pi$. 
The negative solutions are not physical, but give a much better impression
about the general behavior of the solutions. 
}
\label{dprm_func}
\end{figure}
%

%
%
%
\def\BigColSep{\setlength{\arraycolsep}{0pt}}
\renewcommand{\arraystretch}{1.2}
\begin{table*}
{Benchmark point B1} 
\begin{center}
\begingroup\BigColSep
\begin{tabular}{|cc|ccccc|}
\hline
$\tan\beta$ & $\alpha/\pi$ 
 & $m_{h^0}$ & $m_{H^0}$ & $m_{A^0}$ & $m_{H^{\pm}}$ & $m^{2}_{12}$ 
\\
\hline
1.75 & -0.1872 & 125.18 & 300 & 441 & 442 & 38300
\\
\hline
\end{tabular}
\end{center}
\vspace{5pt}
\begin{flushleft}{Higgs potential parameters in the generic basis}\end{flushleft} 
\begin{center}
\begin{tabular}{|ccc|ccccccc|}
\hline
$m_{11}^{2}$ & $m_{22}^{2}$ & $m_{12}^{2}$ 
& $\lambda_{1}$ & $\lambda_{2}$ & $\lambda_{3}$
  & $\lambda_{4}$ & $\lambda_{5}$ & $\lambda_{6}$ & $\lambda_{7}$
\\
\hline
{\small 63484} & {\small 12414.5} & {\small 38300} 
& {\small 0.00653748} & {\small 0.36458} & {\small 3.66474}
 & {\small -1.77052} & {\small -1.74139} & {\small 0} & {\small 0}
\\
\hline
\end{tabular}
\end{center} 
\vspace{5pt}
\begin{flushleft}{Higgs potential parameters in the Higgs basis}\end{flushleft} 
\begin{center}
\begin{tabular}{|ccccccc|}
\hline
& $Y_{1}$ && $Y_{2} $ && $Y_{3}$ & 
\\
\hline
& -8011.44 && 83910. && -2554.55 &
\\
\hline
\hline
$Z_{1}$ & $Z_{2}$ & $Z_{3}$
  & $Z_{4}$ & $Z_{5}$ & $Z_{6}$ & $Z_{7}$
\\
\hline
0.264299 & 0.082535 & 3.67689
  & -1.75837 & -1.72924 & 0.0842752 & 0.0699585
\\
\hline
\end{tabular}
\endgroup
\end{center}
\caption{Benchmark point B1 of~\cite{Hespel:2014sla}. The value of the
lightest Higgs boson $h^{0}$ is updated to the newest PDG 
value~\cite{PDG2018}. The vacuum expectation value 
$v^{2} = G_{F}^{-1} / \sqrt{2}$, needed to calculate the potential 
parameters in the generic or in the Higgs basis, is defined in the same way 
as in~\cite{Hespel:2014sla}, but the value for 
$G_{F} = 1.1663787(6)\times10^{-5}\,\text{GeV}^{-2}$ is taken
from~\cite{PDG2018}. The bilinear parameters $m_{jk}^{2}$ or $Y_{i}$ are 
given in GeV$^{2}$.
} \label{table3}
\end{table*}
\renewcommand{\arraystretch}{1.0}

To better illustrate the behavior of the solutions of the 4th order equation, 
eq.~(\ref{4o-eq-dp}), we pick the benchmark point B1 
of~\cite{Hespel:2014sla}, summarized in table~\ref{table3},
and show the solutions in the 
$|d^{\prime}|$-$\phi^{\prime}$ plane
as lines with different values $\lambda_{D}$ for the fixed 
Majorana mass $M_{R} \sim m_{4} = 10^{10}\,$GeV in
figure~\ref{dprm_func}.
There are 4 solutions for each value of $\phi^{\prime}$, but only real
solutions $|d^{\prime}|$ are displayed.
The solutions are indexed by their algebraic expressions ($1^{st}$, $2^{nd}$,
$3^{rd}$, and $4^{th}$). Their order is not related to their magnitude or
the numeric nature (whether the value is real or complex).
The left panel of figure~\ref{dprm_func} shows the solutions 
for a large variation in $\lambda_{D}$. When $\lambda_{D} = 0.5$ or 
$\lambda_{D} = 1$, we find one physical solution $|d^{\prime}|>0$,
one non-physical solution $|d^{\prime}|<0$, and two complex
$|d^{\prime}|$ solutions for every phase value $\phi^{\prime}$.
But we see also at certain phase values 
$\phi^{\prime} = \pi \pm \pi/2$, that the index numbers
of the solutions switch:
at $\phi^{\prime} = \pi/2$ the $1^{st}$ and $2^{nd}$ solutions become
complex and the earlier complex $3^{rd}$ and $4^{th}$ solutions become 
real. 
When $\lambda_{D} = 1.5$ or $\lambda_{D} = 2$, we find two pairs of real 
solutions, but only in a very limited range of the phase $\phi^{\prime}$. 
These two physically
allowed values of $|d^{\prime}|$, the $3^{rd}$ and $4^{th}$ solutions,
will give different Yukawa couplings,
for the same point in the Higgs sector. The plot also
contains colored dots that are later used to illustrate the value of
the Yukawa couplings.
The right panel of figure~\ref{dprm_func} shows how sensitively the number
of positive solutions can depend on the parameters of the model: 
for $\lambda_{D} = 1.0892$ and 
$\phi^{\prime} \lesssim \pi/2$ we get 
three positive solutions and one negative solution.
Specifically, when 
$\phi' = 3\pi/8$
the real positive solutions are 
$|d'| = 10^{-3} \times \{ 0.00892, 0.3698, 6.5685 \}$.

%
%
%
\begin{figure}[t]
\begin{center}
\includegraphics[width=1.0\textwidth]{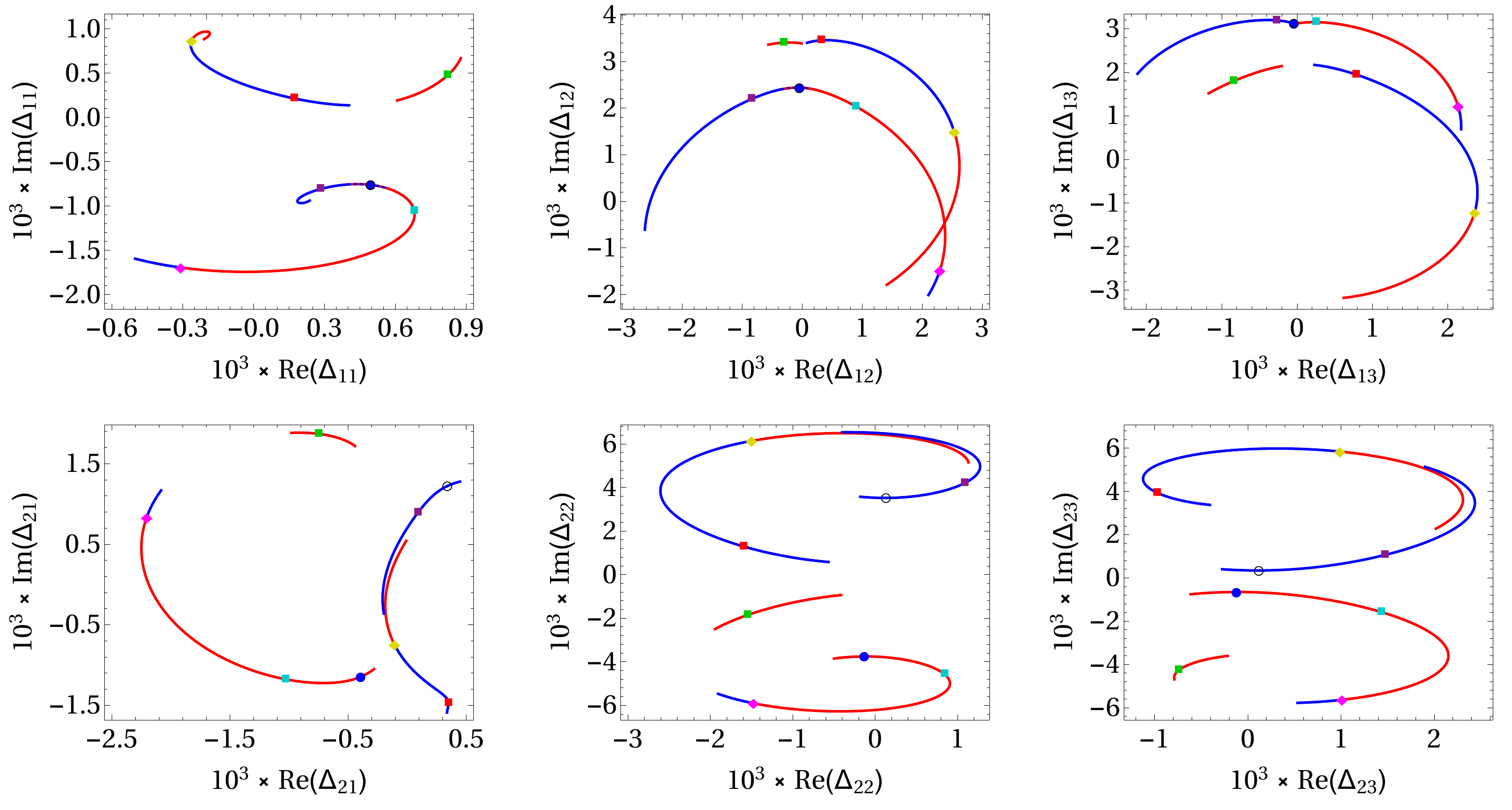}
\vspace{-20pt}
\end{center}
\caption{(Color online) The values of the Yukawa couplings 
$\Delta_{1}$ and $\Delta_{2}$ in the complex plane
for the fixed values of $\lambda_{D}=1$ and
$M_{R} \sim m_{4} = 10^{10}\,$GeV. The Higgs parameters correspond to the
benchmark point B1, table~\ref{table3}. The normal hierarchy is assumed.
The colors of the curves correspond to the colors of the solutions in 
figure~\ref{dprm_func}. 
}
\label{Yuk_func}
\end{figure}

Even though the discussion of the possible values of $d$ and $d^{\prime}$ is 
interesting and not too simple by itself,
it does not show physical observables,
but theoretical constructs. Possible physical observables are the 
Yukawa couplings $\Delta_{1}$ and $\Delta_{2}$, which we show in
figure~\ref{Yuk_func} for the same benchmark point B1, table~\ref{table3}.
Even with fixed values of $M_{R}$
and $\lambda_{D}$ we do not get separate points but curves in the complex plane
for each component of the two Yukawa couplings $\Delta_{1}$ and $\Delta_{2}$.
These curves result from the sum of two different columns of the
PMNS matrix with complex coefficients, eqs.~(\ref{Delta1}) and (\ref{Delta2}).
They can be additionally multivalued in other cases
(different from B1), because we can have two,
three, or four solutions to the fourth order equation for
$|d^{\prime}|$, eq.~(\ref{4o-eq-dp}).

The values of $|d^{\prime}|$, marked in figure~\ref{dprm_func}, lead
to different values of the Yukawa couplings shown in
figure~\ref{Yuk_func}. The blue and red lines in
figure~\ref{dprm_func} mark the values corresponding to the $2^{nd}$
and $4^{th}$ solution, respectively. Those solutions sometimes lead to
identical values of $\Delta_{1k}$, as shown in figure~\ref{Yuk_func},
where the red-blue dashed line is marked by a black open circle that
is positioned on top of a filled blue circle. These two reference points
lead to different values of $\Delta_{2k}$, as shown in the lower plots
of figure~\ref{Yuk_func}.

\begin{figure}[t]
\begin{center}
\includegraphics[width=1.0\textwidth]{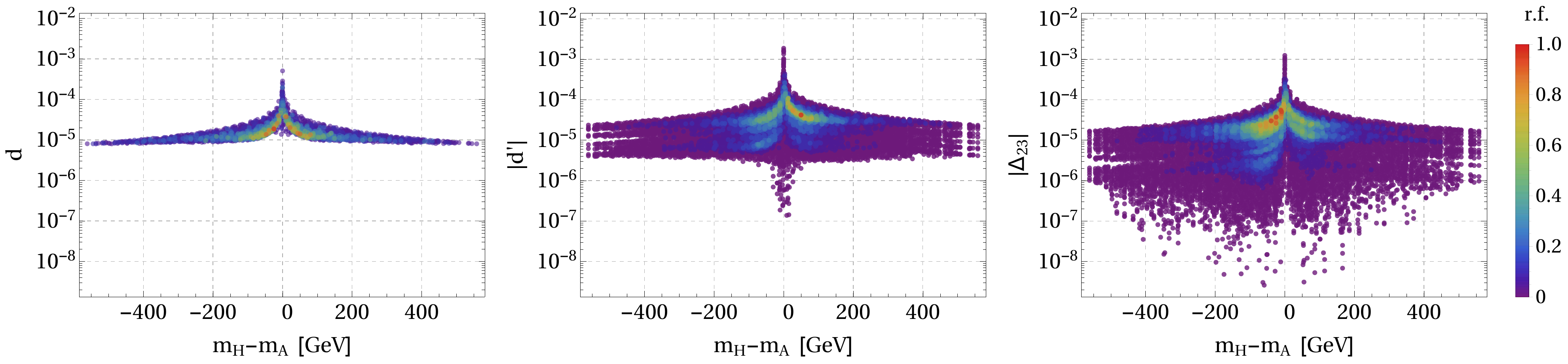}
\vspace{-20pt}
\end{center}
\caption{(Color online) Scatter plot of the values of the parameters
  $d$ and $|d'|$ (left and center) and the Yukawa coupling
  $|\Delta_{23}|$ (right) as a function of the Higgs boson mass
  difference $m_{H}-m_{A}$. We used $m_{4}=10^{5}$~GeV, and
  $\lambda_{D}=1$ for this plot.
  5000 points of the Higgs sector are used to have better statistics.
  The points are colored according to their normalized relative
  frequency (r.f.) of occurrence.
}
\label{Delta23_plots}
\end{figure}

Describing the distributions of parameters $d$ and $|d'|$
(shown in figure~\ref{dpdm4-plot}), we already discussed their
dependence on the Higgs masses. This dependence is illustrated in
figure~\ref{Delta23_plots}, where their values and the
size of the Yukawa coupling $|\Delta_{23}|$ are plotted as a function
of $(m_{H}-m_{A})$ for $m_{4}=10^5$~GeV and $\lambda_{D}=1$.
The values of $d$ have a clear lower bound and get larger when
$m_{H}$ gets closer to $m_{A}$. The values of $|d'|$ are scattered in
a wider range, and the values can get smaller or larger, when
$m_{H}\approx m_{A}$. The values of the Yukawa coupling
$|\Delta_{23}|$ are more scattered.
5000 points of the Higgs sector were used for the plot,
which resulted in 5000 values of $d$, around 110.000 values of $|d'|$,
and 150.000 values of $|\Delta_{23}|$.
The number of values for $|d'|$ and $|\Delta_{23}|$ depends on sampling
algorithm, because they depend on a free parameter $\phi'$.
The scattered values are colored
according to the relative frequency of value occurrence.
The difference $m_{H}-m_{A}<100$~GeV is dominating, because the
applied restrictions on the Higgs sector lead to $m_{H}$ getting close
to $m_{A}$ as their masses increase, and we equalized the distribution
of points in the $(m_{H},m_{A})$ plane.

%
%
%
%
%

%
%
%
\begin{figure}[t]
\begin{center}
\includegraphics[width=1.0\textwidth]{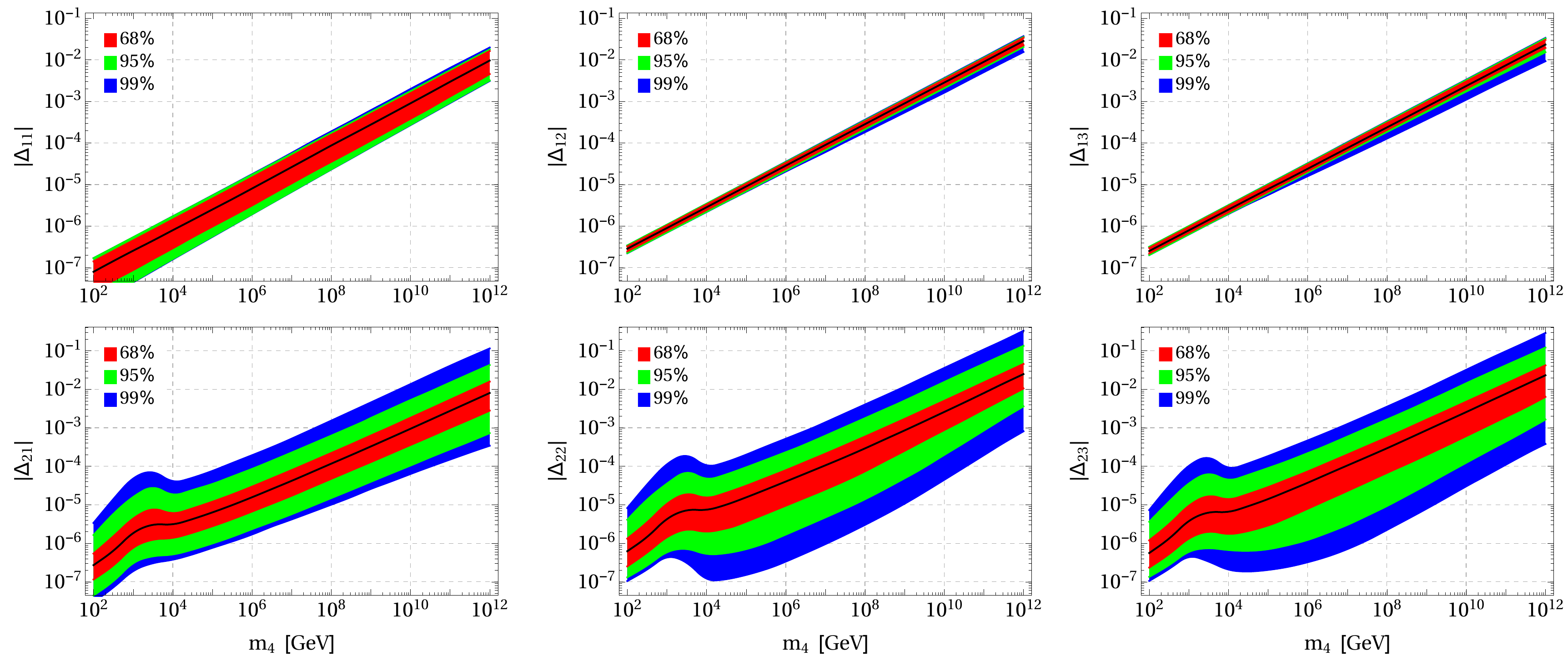}
\vspace{-20pt}
\end{center}
\caption{(Color online) The modulus of the values of the Yukawa couplings 
$\Delta_{1}$ and $\Delta_{2}$ as functions of $M_{R} \sim m_{4}$
for the fixed value of $\lambda_{D}=1$ for the normal hierarchy. 
For each value of $m_{4}$ we show over 32.000 points that come from 
varying the phase $\phi^{\prime}$ for each of the 1.000 points in the Higgs 
sector, like in figure~\ref{dpdm4-plot}. 
The black line marks the median of these 32.000 values for
each value of $m_{4}$. The 68\% of values of $|\Delta_{jk}|$
closest to the median are shown in red, the values in the range of
68\% to 95\% are shown in green,
and the values in the range of 95\% to 99\% are shown in blue.
We do not show the
values outside the range of 99\%, as they would fill up the rest of the 
plot and no information could be obtained by looking at it.
}
\label{Yuk_m4}
\end{figure}
%
%
%
%
\begin{figure}[t]
\begin{center}
\includegraphics[width=1.0\textwidth]{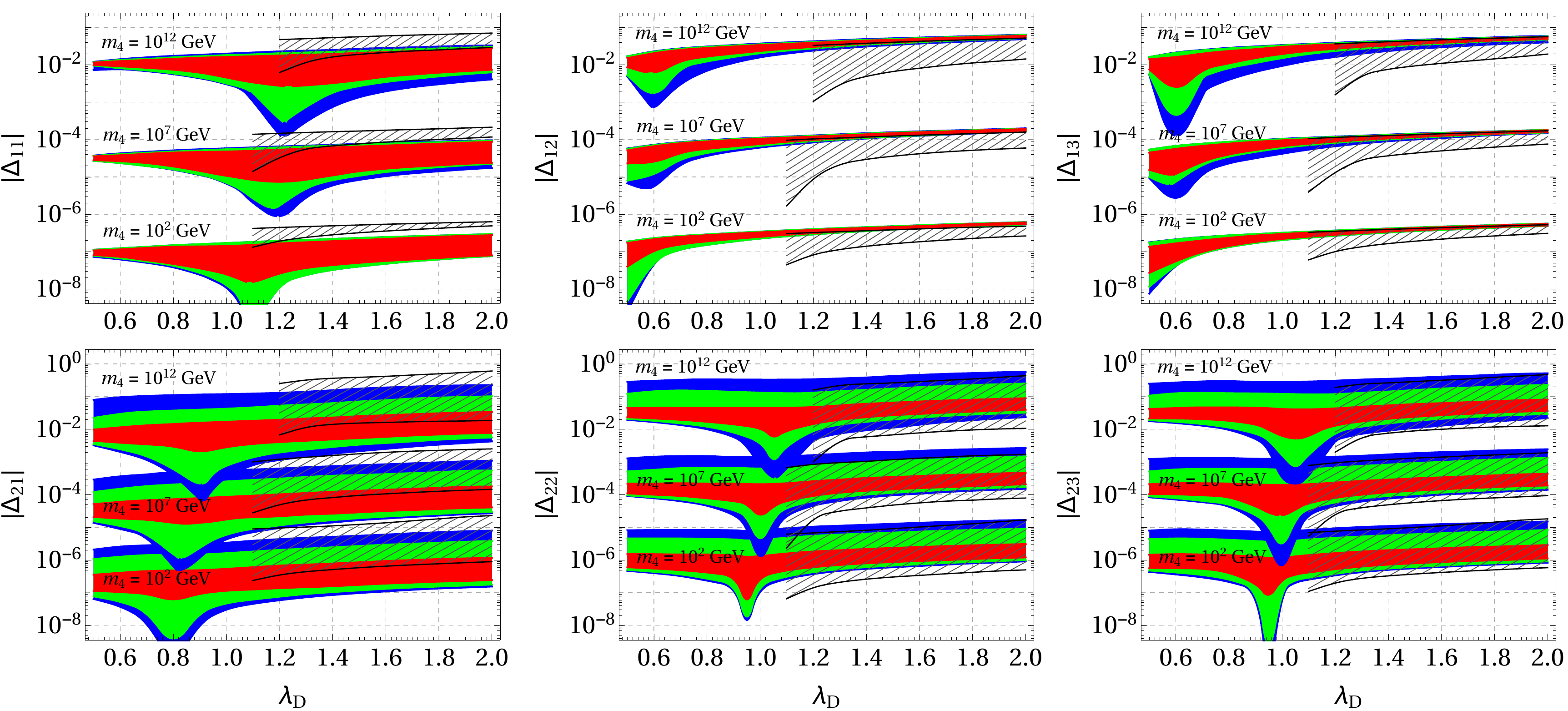}
\vspace{-20pt}
\end{center}
\caption{(Color online) The modulus of the values of the Yukawa couplings
$\Delta_{1}$ and $\Delta_{2}$ as functions of $\lambda_{D}$
for the fixed values of $m_{4} = \{ 10^{2}, 10^{7}, 10^{12} \}\,$GeV. 
The color coding of the ranges is the same as in figure~\ref{Yuk_m4}, 
but we do not show the median. Additionally, we show the range of 99\% of the
values of $|\Delta_{jk}|$ for the inverted
hierarchy by the striped area. This striped area starts with the values
of $\lambda_{D} = 1.1$ or $\lambda_{D} = 1.2$ because for smaller 
values of $\lambda_{D}$ we do not get enough solutions to derive reliable
statistics.
}
\label{Yuk_lD}
\end{figure}
%
%
%
%
\begin{figure}[t]
\begin{center}
\includegraphics[width=0.6\textwidth]{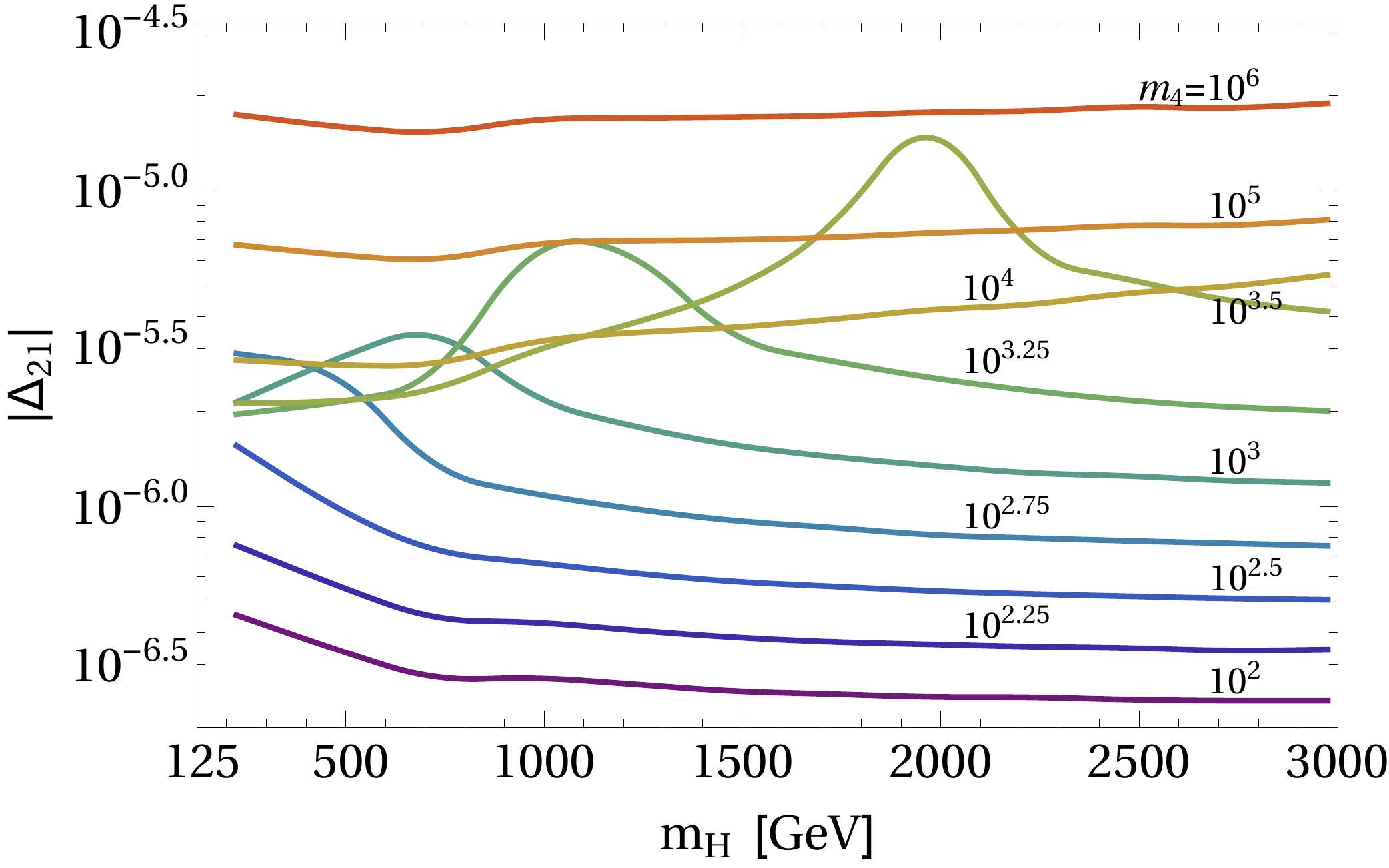}
\vspace{-20pt}
\end{center}
\caption{(Color online) 
The median of $|\Delta_{21}|$
as a function of $m_{H}$ for the fixed values of $m_{4}$ as
displayed in the plot close to each line.
The values were calculated in intervals of $\Delta m_{H}=250$\,GeV and
splined.
$\lambda_{D}=1$ is fixed for all curves. Each line
contains the statistical information from more than 150.000 parameter 
points that are taken in the respective range of $m_{H}$, but varying 
over $m_{A}$, $(\beta-\alpha)$, and the allowed values of $\phi^{\prime}$.
}
\label{Yuk_mH}
\end{figure}

We continue our discussion with the
statistical description of the whole parameter space of our model. In 
figure~\ref{Yuk_m4} we show the distribution of the size of individual 
components of the Yukawa couplings for $\lambda_{D}=1$ fixed and $m_{4}$
varying between $10^{2}$ to $10^{12}$ on a logarithmic scale. 
The components of the first Yukawa coupling $\Delta_{1}$ show only
little variation
and a linear dependence in the logarithmic plot on $m_{4}$. But the 
second Yukawa coupling $\Delta_{2}$ exhibits a much larger variation 
and also a structure at low values of $m_{4}$. This comes from the 
definition of the Yukawa couplings, eq.~(\ref{Delta1}) and (\ref{Delta2}). 
Whereas $\Delta_{2}$ carries the whole variation of $d$ and $d^{\prime}$, 
$\Delta_{1}$ only sees the dependence of $R_{3}$, eq.~(\ref{R-def}), 
and of $m_{D}$, eq.~(\ref{def_scaling}). 

In figure~\ref{Yuk_lD} we see the distribution of the size of individual 
components of the Yukawa couplings $|\Delta_{jk}|$
for three values of $m_{4}$ in dependence on $\lambda_{D}$.
The smooth upper value for the $|\Delta_{jk}|$ corresponds to the upper edge
of the distributions of $d$ and $|d^{\prime}|$ in figure~\ref{dpdm4-plot}.
The running dips of the lower values of the $|\Delta_{jk}|$ can be
understood by the fact that the $\Delta_{jk}$ are sums of complex numbers
that depend smoothly on the parameter $\lambda_{D}$.
Namely, the variation over
the phase $\phi^{\prime}$ can give one very small Yukawa coupling 
$\Delta_{jk}$, as seen in figure~\ref{Yuk_func}. Together with the 
upper value for $d$ and $|d^{\prime}|$ (see fig.~\ref{dpdm4-plot})
the variations over the phase $\phi^{\prime}$ and over the points
of the Higgs potential for a given $\lambda_{D}$ produce
the larger spread of values $|\Delta_{jk}|$, seen as the dips in 
figure~\ref{Yuk_lD}. 

The striped regions in figure~\ref{Yuk_lD} depict the values of 
$|\Delta_{jk}|$ for the inverted hierarchy. One can notice the similarity 
of the vertical thickness
of (a)~the striped areas for $|\Delta_{j2}|$ and $|\Delta_{j3}|$
with the colored area of $|\Delta_{j1}|$, and (b)~the striped area of
$|\Delta_{j1}|$ with the colored areas of $|\Delta_{j2}|$ and
$|\Delta_{j3}|$. The behavior reflects the exchange of the related
neutrino states: in the inverted hierarchy the two heavier states are 
more similar whereas in the normal hierarchy the two lighter 
states are closer related. In some way $|\Delta_{11}|$ represents the 
decoupled state for the normal hierarchy and by that the vector of
the PMNS matrix that stands for the massless neutrino. $|\Delta_{12}|$ and 
$|\Delta_{13}|$ give the states mixed from the seesaw mechanism and 
radiative mass generation. 
In the inverted hierarchy it is $|\Delta_{13}|$
that represents the massless neutrino and $|\Delta_{11}|$ and 
$|\Delta_{12}|$ that give the mixed states.
This behavior is not so pronounced in $\Delta_{2k}$, since this 
Yukawa coupling is the superposition of the PMNS vectors with the complex 
numbers $d$ and $d^{\prime}$, giving a much larger spread of values, 
as could already be seen in figure~\ref{Yuk_m4}.

Figure~\ref{Yuk_mH} shows the wave-like behaviour of the 
median of $|\Delta_{21}|$ that comes from 
the interplay between the scale of the Higgs boson masses and the 
scale of the Majorana mass term. A hint for this interesting behavior is 
already seen in the bump of $|\Delta_{21}|$ for low values of $m_{4}$ 
in figure~\ref{Yuk_m4}. We obtain very similar plots for $|\Delta_{22}|$
and $|\Delta_{23}|$, as both elements of the second Yukawa coupling
have a similar dependence on the parameters of the model.

\subsection{Numerical advantage of the analytic approach}
\label{Num:advantage}

In our analysis we can find observables which satisfy the experimental bounds 
by scanning over only one parameter, for example the phase
$\phi^{\prime}$. But the usual way for the calculation of observables 
in such a model (or more sophisticated models) is fitting the parameters 
by using some global minimization algorithm. 
Due to the small number of parameters and observables in our study
there is a good
possibility to compare these two different methods of calculation. 
In order to find the numerical values for the parameters we construct a
minimization function $\chi^2$ 
\begin{equation}
\chi^2 
= 
\sum_{i=1}^{n}
\left( 
  H \left(O_i^v - \bar{O_i}\right) 
  \left( \frac{O_i^v - \bar{O_i}}{\delta_+ O_i}\right)^2 
+ H \left(\bar{O_i} - O_i^v\right) 
    \left( \frac{\bar{O_i} - O_i^v}{\delta_- O_i}\right)^2 
\right)
\enspace ,
\label{chi2}
\end{equation}
where $n$ is the number of observables to be fitted. In this case we fit 
the neutrino masses and the oscillation parameters $\theta_{12}$, 
$\theta_{13}$, $\theta_{23}$, and $\delta_{CP}$. $H$ is the Heaviside 
step function, $\bar{O_i}$ denotes the central value of each observable $O_i$, 
$\delta_{\pm} O_i$ are the upper and lower experimental errors of the
observable,
and $O_i^v$ is the calculated value of the observable. The data is fitted by 
minimizing $\chi^2$ with respect to the Yukawa couplings $\Delta_1$ and 
$\Delta_2$ in eq.~(\ref{Yukawa}), which means that there are twelve 
real parameters to be fitted. The central values 
(i.e.\ the experimental best fits) $\bar{O_i}$ and the $1\sigma$ errors
$\delta_{\pm} O_i$
are taken from~\cite{deSalas:2017kay}. 
%
%
\begin{figure}[t]
\begin{center}
\includegraphics[width=0.6\textwidth]{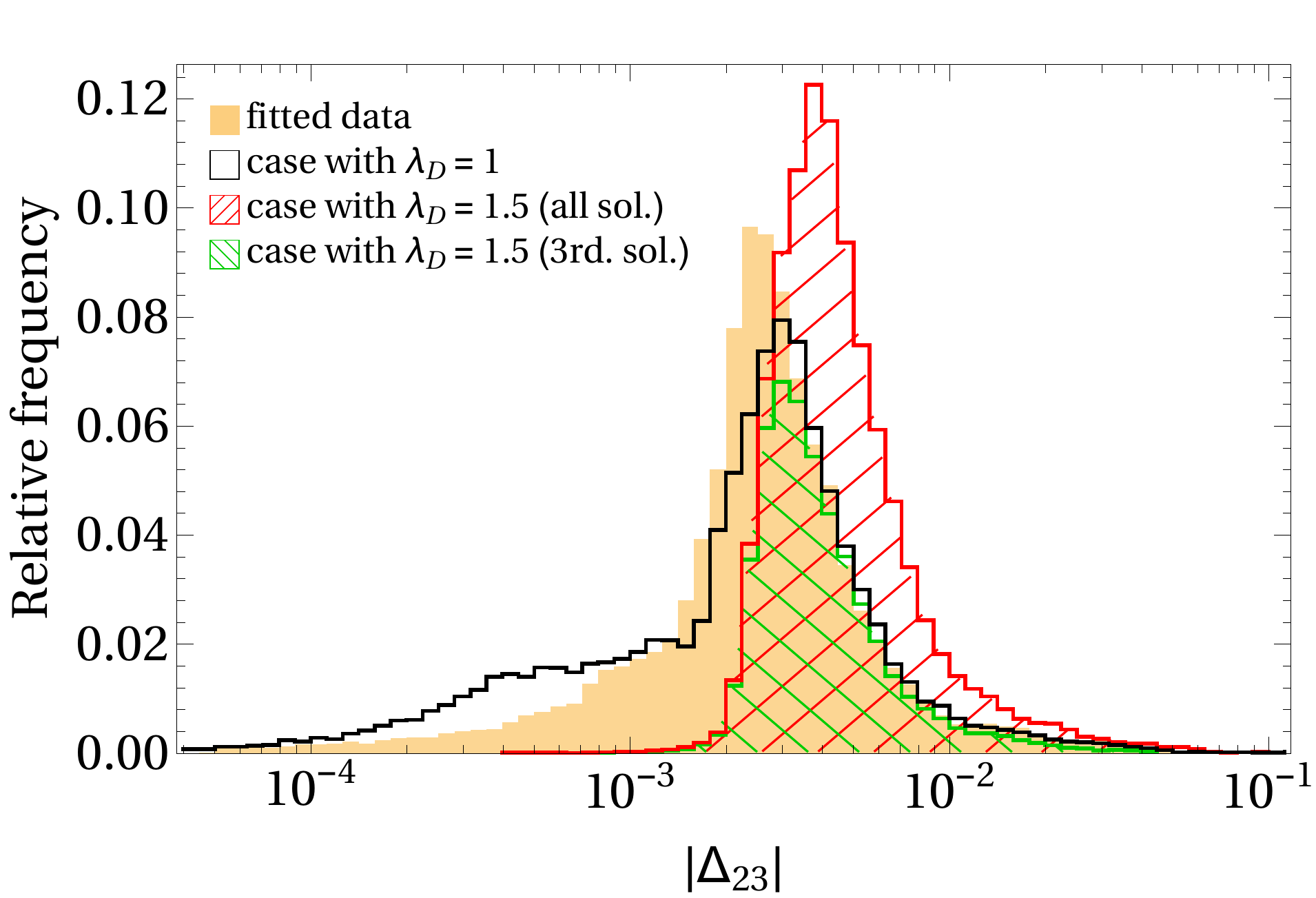}
\vspace{-20pt}
\end{center}
\caption{(Color online) Comparison of the Yukawa coupling $\Delta_{23}$ 
obtained with different methods. The relative frequency of the value of
$|\Delta_{23}|$ is obtained by counting how often the value appears in the 
selected bin, divided by the total number of points. 
We assume the normal hierarchy and take $M_{R} \sim m_{4} = 10^{10}\,$GeV fixed.
The yellow area represents data points obtained with the minimization algorithm, 
while the black and red histograms represent the distributions of 
$|\Delta_{23}|$ calculated with our analytical procedure, described in 
sec.~\ref{framework:improved couplings}. 
The black histogram uses $\lambda_D = 1$ and the red histogram 
$\lambda_D = 1.5$. The green histogram shows a part of the red
histogram that is determined by taking only
the third solution of the fourth order equation~(\ref{4o-eq-dp}).
}
\label{picture_fits}
\end{figure}

For the numerical minimization of the $\chi^2$ function we have used the 
``differential evolution'' algorithm which is expensive with respect to
computer resources but quite effective. The calculations were carried out 
for the 1.000 points in the Higgs sector that were used also for 
Fig.~\ref{dpdm4-plot}, by running forty to sixty separate minimizations 
on each data point.\footnote{Sometimes the minimization algorithm does 
not find the global minimum to the desired precision. Therefore other 
attempts are performed until at least forty minimum points are collected. 
We limit our tries to sixty attempts. Most of the time fifty attempts 
are enough to find forty converging minima points.}
The parameter sets were saved 
if $\chi^2 < 10^{-15}$, meaning that the calculated value of the observables 
coincide to a high accuracy with the respective experimental central value.  

In figure~\ref{picture_fits} we compare the distribution of the Yukawa coupling  
$\Delta_{23}$ calculated using different methods. The histogram in the 
figure shows the statistical distribution of more than 40.000 points. 
The yellow area represents data obtained by the minimization algorithm while 
the black and red histograms represent the distributions of $\Delta_{23}$ 
calculated with the analytical procedure, described in 
sec.~\ref{framework:improved couplings}, for $\lambda_D = 1$ or 
$\lambda_D = 1.5$, respectively. We see that the black histogram almost 
coincides with the fitted data, but the red histogram is moved to larger 
values of $|\Delta_{23}|$. As can be seen in Fig.~\ref{dprm_func}, 
$\lambda_D = 1.5$ restricts the phase $\phi'$ to a rather small interval, 
but gives larger values of $|d'|$ at the same time. Since $|\Delta_{23}|$
depends strongly on $|d'|$, the larger value of $|d'|$ explains the 
shift in the distributions. But the minimization algorithm just looks for 
any solution and therefore finds the points more probably in larger areas 
of the parameter space. Since $\lambda_D = 1$ has a larger phase space 
than $\lambda_D = 1.5$, the distribution coming from the minimization algorithm 
should be more similar to the histogram with $\lambda_D = 1$, which is 
what we see in figure~\ref{picture_fits}.
For $\lambda_D = 1.5$ approximately half of the values of $\Delta_{23}$ come 
from the third solution of the fourth order equation~(\ref{4o-eq-dp}). These
values are shown by the green histogram in figure~\ref{picture_fits}. 
Approximately another half of the values come from the fourth solution 
and only a very small fraction of values is given by other solutions. 

From this example we can guess that using the minimization algorithm in 
the general case, i.e.\ by varying twelve free parameters, we could miss some
regions of the parameter space, if we do not repeat the minimization often 
enough. But the main difference between the fitting 
procedure and calculations using our analytical method 
is the usage of computational resources. The calculations described in this 
example took about 430 times longer using the minimization method 
than using our analytical method.

\section{Summary} \label{summary}

The seesaw mechanism is one of the most successful extensions of the
SM which explains neutrino masses. In the usual setup, one adds a heavy 
singlet fermion for each light neutrino. The Grimus-Neufeld model adds 
only a single Majorana fermion to the fermion content of the SM, producing 
only a single seesaw mass for the SM-like neutrinos.
Finite corrections to the neutrino 
mass matrix arise from one-loop diagrams mediated by the heavy neutrino. 
In the Grimus-Neufeld model these loop corrections produce a radiative mass 
for one SM-like neutrino. In order to allow this radiative mass
the Higgs sector of the Grimus-Neufeld model is constructed from two Higgs
doublets, giving two Yukawa couplings to the heavy neutrino. These 
Yukawa couplings have to be linearly independent, thus characterizing
the Higgs sector of the model as a general type. 
For simplicity we assume a CP-invariant Higgs potential. 
For the numerical calculations, 
we take the masses of the neutral Higgs bosons and their mixing angle 
as input parameters. 

We parameterize the Yukawa couplings to the heavy neutrino and calculate 
the neutrino masses and oscillation parameters following the approximations
of Grimus and Lavoura ~\cite{Grimus:2002nk}. 
Since we obtain analytical solutions for the neutrino masses, and the
Grimus-Neufeld model has the lightest neutrino massless at one loop 
level~\cite{Dudenas:2018wlr}, we can use the two measured mass differences
as input to determine the Yukawa couplings. With this approach we also
retain the neutrino mixing matrix as an unchanged input for our calculation. 
This change in the parameterization is a new feature compared to 
previous treatments of seesaw models and has the major advantage, that 
it reduces the undetermined parameters of the model.

After the distribution of tree-level heavy Higgs masses and the physical 
tree-level mixing angle between $h^0$ and $H^{0}$ in figure~\ref{higgses}
we show the distribution of the parameters $d$ and $|d^{\prime}|$, 
that parameterize the second Yukawa coupling, in 
dependence of the heavy Majorana mass $m_{4}$ in figure~\ref{dpdm4-plot}. 
Taking the benchmark point B1 from~\cite{Hespel:2014sla} as a reference,
we show in figure~\ref{dprm_func} the behavior of the solutions of the
fourth order equation that we need to solve to obtain $|d^{\prime}|$
and in figure~\ref{Yuk_func} we show the corresponding Yukawa couplings. 

We present a statistical analysis of the modulus of the Yukawa couplings 
in dependence on $m_{4}$ in figure~\ref{Yuk_m4} and in dependence
on $\lambda_{D}$ in figure~\ref{Yuk_lD}. As a final plot~\ref{Yuk_mH} 
in the presentation of the parameter space
we also show the wave of the median of $|\Delta_{21}|$ depending 
on the mass of the heavy scalar Higgs and discuss its origin, 
finishing the overview over the parameter space of the 
Grimus-Neufeld model. 
The last subsection illustrates with figure~\ref{picture_fits} 
the numerical advantage of finding 
an analytical solution. 

In summary, we parameterized and discussed the Grimus-Neufeld model in
terms of mostly physically measured low energy scale quantities. The
only two ``non-physical'' parameters that are used in our model are
(1)~the phase of the Yukawa coupling $d'$ of the tree-level ``seesaw''
neutrino mass state to the second Higgs doublet, denoted as $\phi'$,
and (2)~the proportionality between the tree-level ``seesaw'' neutrino
mass and its mass after the 1-loop radiative correction, denoted as
$\lambda_{D}$. The other parameters are directly measureable quantities. 
Having only two not directly measureable parameters increases the 
testability of our model: a few measurements that restrict the 
neutrino Yukawa couplings can confirm or rule out our model. 

Our study of the Grimus-Neufeld model does not end here. This paper
discussed only the Higgs and neutrino sectors.  We aim to study the
full model with all particle sectors included and get additional
restrictions on the Grimus-Neufeld model parameter space from the
estimated predictions of rare processes.


\begin{acknowledgments}
The authors thank the Lithuanian Academy of Sciences for the support.
\end{acknowledgments}

\appendix

\section{Neutral Higgs mass eigenfields}
\label{Appendix-b-vectors}

Some features of the formalism for the scalar sector of a
multi-Higgs-doublet SM are given in
ref.~\cite{Grimus:1989pu,Grimus:2002nk,Grimus:2002prd}. Here we discuss the
properties of the vectors $b$ and give expressions for their
calculation in the case of two Higgs doublets.

The physical neutral scalar mass eigenfields are expressed as
\begin{equation}\label{sumb-appendix}
\phi_{b_k}^0=\sqrt{2} \sum_{j=1}^{n_H} \mathrm{Re}(b_{kj}^* \phi_j^0)
= \frac{1}{\sqrt{2}} \sum_{j=1}^{n_{H}}
 \left( b^{*}_{kj} \phi^{0}_j +
   b_{kj} \phi^{0\,*}_j \right),
\end{equation}
which are characterized by $2 n_H$ unit vectors $b_k \in
\mathbbm{C}^{n_H}$ of dimensions $n_H \times 1$.
In the matrix-vector
notation, these eigenfields can be written as
$\phi_{b_k}^0=\sqrt{2}\,\mathrm{Re}(b_k^{\dagger} \phi^0)$.

The orthonormality equations for the vectors are
\begin{eqnarray}
&&\sum\limits_{j=1}^{n_H} \left(\mathrm{Re}(b_{kj})
 \mathrm{Re}(b_{k^{\prime}j}) + \mathrm{Im}(b_{kj})
 \mathrm{Im}(b_{k^{\prime}j})\right)
=
\sum\limits_{j=1}^{n_H} \mathrm{Re}(b_{kj}^* b_{k^{\prime}j})
  =\delta_{b_k b_{k^{\prime}}};
  \label{orthog1a}
\\
&& \sum\limits_{k=1}^{2n_H} \mathrm{Re}(b_{kj}) \mathrm{Re}(b_{k j^{\prime}})
  =\sum\limits_{k=1}^{2n_H} \mathrm{Im}(b_{kj}) \mathrm{Im}(b_{k j^{\prime}})
  =\delta_{jj'};
  \label{orthog1b}
\\
&& \sum\limits_{k=1}^{2n_H} \mathrm{Re}(b_{kj}) \mathrm{Im}(b_{k j^{\prime}})
 =\sum\limits_{k=1}^{2n_H} b_{kj} b_{k j^{\prime}}=0.
 \label{orthog1c}
\end{eqnarray}
The vectors $b_k$ and $b_{k^{\prime}}$ indicate two different states
$\phi_{b_k}^0$ and $\phi_{b_{k^{\prime}}}^0$, and indices $j$ and $j^{\prime}$
indicate two different components of the vectors $b$.

The neutral Goldstone boson $G^0 = \phi^0_{G^0}$
corresponds to the vector $b_{G^0}$ with the components
$\left( b_{G^0} \right)_j = i v_j / v$ \cite{Grimus:1989pu,Grimus:2002nk,Grimus:2002prd},
where
$v = \left( | v_1 |^2 + | v_2 |^2
+ \cdots + | v_{n_H} |^2 \right)^{1/2}
= 2 m_W / g$.
In the case of only two Higgs doublets, and due to the rotation of the
Higgs fields to make the vacuum expectation value a feature of the
SM Higgs field, the vector $b_{G^0}$ equals
\begin{equation}\label{bG0}
b_{G^0}= \left(\begin{array}{c} i \\ 0 \end{array}\right) \, .
\end{equation}

Physical Higgs fields $\phi^0_{b_k \neq G^0}$ must be orthogonal to
the Goldstone field
$G^0$
which follows from \refeq{orthog1a}. This leads to the condition
\begin{equation}
\sum_{j=1}^{n_H} \mathrm{Re} \left( -\frac{i v_j}{v} b^*_{kj} \right) =
 \frac{1}{v} \sum_{j=1}^{n_H} \mathrm{Im} \left( v_j b^*_{kj} \right) =
 \sum_{j=1}^{n_H} \mathrm{Re} \left( b_{G^0j} \, b^*_{kj} \right)=0.
\label{orthog2-append}
\end{equation}

To study the unit vectors $b$, introduced in eq.~\refeq{sumb-appendix}
(which are the same as eq.~\refeq{sumb} in the text) and corresponding to
the Higgs fields other than the Goldstone boson $G^0$, lets define
them in the following form:
\begin{equation}
b_1 = \left( \begin{array}{c} b_{11} \\ b_{12} \end{array} \right) ,
\qquad
b_2 = \left( \begin{array}{c} b_{21} \\ b_{22} \end{array} \right),
\qquad
b_3 = \left( \begin{array}{c} b_{31} \\ b_{32} \end{array} \right).
\label{bVectSetGeneral}
\end{equation}
From the orthogonality relations (\ref{orthog1a} - \ref{orthog1c}) and
due to the fixed value of $b_{G^0}$~\refeq{bG0} it is possible to write the
orthogonality equations for the vector components in the following manner:
\begin{eqnarray}
&&
 b_{11}, b_{21}, b_{31} \in \mathbbm{R};
 \qquad
 b_{12}, b_{22}, b_{32} \in \mathbbm{C};
 \label{beq1a}
\\
&&
 b_{k1}^{2} + \left| b_{k2} \right|^2 = 1;
 \label{beq1b}
\\
&&
 b_{k1} b_{k^{\prime}1}
 +\mathrm{Re}\left( b_{k2}^* b_{k^{\prime}2} \right)
 =0;
\\
&&
 \sum_{k=1}^{3} b^2_{k2} = \sum_{k=1}^{3} b_{k1} b_{k2} = 0;
\\
&&
 \sum_{k=1}^{3} b^{2}_{k1} =
 \sum_{k=1}^{3} \left[\mathrm{Re} \left(b_{k2} \right)\right]^2
 = \sum_{k=1}^{3} \left[\mathrm{Im} \left(b_{k2} \right)\right]^2 = 1.
 \label{beq1d}
\end{eqnarray}

By choosing $b_{21}$, $b_{31}$, and $\mathrm{Re} ( b_{32} )$ as
input variables,
it is possible to express the other
components of the vectors $b$ by those variables by solving the
equations (\ref{beq1a} - \ref{beq1d}). Introducing three
sign-parameters $s_{32\mathrm{im}}$, $s_{11}$, and $s_{22}$ (they can take
values $\pm1$), we can write
\begin{align}
 \mathrm{Im} \left( b_{32}\right)=&
 s_{32\mathrm{im}} \sqrt{1- b_{31}^2-
   \left[\mathrm{Re} \left( b_{32}\right)\right]^2}\enspace;
 \label{beq2a}
\\
 b_{11}=& s_{11} \sqrt{1- b_{31}^2-b_{21}^2}\enspace;
 \label{beq2b}
\\
 b_{\mathrm{comb}}\equiv&
   \frac{b_{31} b_{21} \mathrm{Re} \left( b_{32}\right)
     +
     s_{22} \left | b_{11} \right|
       \left | \mathrm{Im} \left( b_{32}\right) \right|
   }
    {b_{31}^2 -1}\enspace
    ;
\\
 p_{22} \equiv& \left\{
\begin{aligned}
  - \mathrm{Sg}(b_{31}) \mathrm{Sg}(b_{21})
    \mathrm{Sg}(\mathrm{Im} \left( b_{32}\right)),&\
   \mathrm{if}\
    \left |\mathrm{Re} \left( b_{32}\right) \right|
       \leqslant \sqrt{\frac{b_{31}^2 b_{21}^2}{1-b_{21}^2}}\ ,
       \\
   s_{22}\mathrm{Sg}(\mathrm{Re}
    \left(b_{32}\right)) \mathrm{Sg}(\mathrm{Im} \left(
    b_{32}\right)),&\
     \mathrm{otherwise}\ ;
\end{aligned}
\right.
\\
 b_{22} =& b_{\mathrm{comb}} + i p_{22}
   \sqrt{1- b_{21}^2- b^2_{\mathrm{comb}}}\ ,
\label{beq2d}
\\
 b_{12}=& -\frac{1}{b_{11}}
 \left( b_{31} b_{32} + b_{21} b_{22} \right) .
 \label{beq2c}
\end{align}
We introduced two intermediate parameters $b_{\mathrm{comb}}$ and
$p_{22}$, and $\mathrm{Sg}(x)$ is the sign function
\begin{equation}
\mathrm{Sg}(x) =
\left\{ \begin{array}{c} -1,\hspace{0.2cm} x<0 \\ \phantom{-}
 1, \hspace{0.2cm} x \geqslant 0 \end{array}\right.
\enspace .
\end{equation}
It is worth mentioning that the solutions for the parameter values,
given by the equations (\ref{beq2a} - \ref{beq2d}), were obtained
assuming $b_{21}, b_{31} \neq \pm 1$. According to the orthogonality
relations (\ref{beq1a} - \ref{beq1d}) the free scale parameters vary
in the following ranges: $|b_{31}|<1$, $|b_{21}|<\sqrt{1-b_{31}^2}$, and
$|\mathrm{Re}(b_{32})|\leq\sqrt{1-b_{31}^2}$. The
extreme values of $\pm1$ for the parameters $b_{21}$ and $b_{31}$
could be obtained by
the index permutation of the vectors $b_k$ (for example,
$b_{21}=1$ can be obtained by
swapping the values of $b_{11}=1$ and $b_{12}$ with those of
$b_{21}$ and $b_{22}$).

Equations (\ref{beq2a} - \ref{beq2d}) give 8 different solutions for
the vectors $b$, corresponding to two possible values of the
sign-parameters $s_x$ ($x=32\mathrm{im}$, $11$, and $22$). 

The expressions of eqs.~(\ref{beq2a} - \ref{beq2d}) are significantly
simpler, if some input parameters are equal to zero. This can lead to
further simplifications after introducing trigonometric functions.
Let us study the case, when $\mathrm{Re} \left( b_{32}\right) = 0$.
Defining $b_{31} = \sin(\vartheta_{13})$,
$b_{21} = \sin(\vartheta_{12})\cos(\vartheta_{13})$, and taking
$s_{32\mathrm{im}}=s_{11}=1$ but $s_{22}=-1$,
we obtain the following parametric values of the vectors $b$:
\begin{equation}
b_{G^0} = \left( \begin{array}{c} i \\ 0 \end{array} \right),\
b_1 = \left( \begin{array}{c}
   \mr c_{12} \mr c_{13} \\ -\mr s_{12} - i \mr c_{12} \mr s_{13}
   \end{array} \right) ,\
b_2 = \left( \begin{array}{c}
    \mr s_{12} \mr c_{13} \\ \mr c_{12} - i \mr s_{12} \mr s_{13}
    \end{array} \right),\
b_3 = \left( \begin{array}{c}
    \mr s_{13} \\ i \mr c_{13}
   \end{array} \right),
\label{bVectSetTh-appendix}
\end{equation}
where $\mr c_{ij} \equiv \cos(\vartheta_{ij})$ and $\mr s_{ij} \equiv
\sin(\vartheta_{ij})$.

\section{Parameterization of the mixing matrix}
\label{Appendix-oscillation-angles}

Neutrino oscillation angles are introduced using the
neutrino mass diagonalization matrix $U$~\refeq{Mtotal} and
factorizing it to contain the ordinary Pontecorvo-Maki-Nakagawa-Sakata
(PMNS) neutrino mixing matrix~\cite{PDG2018}. We introduce the formalism by
discussing the $3\times 3$ neutrino mixing case, where the
relationships are simpler; then we expand it to the $4\times 4$ case.

The simplest case ($3\times3$) considers only the light neutrinos,
assuming they are Majorana particles. This case is
discussed in ref.~\cite{Dziewit:2011pd} in a slightly different
notation of the matrix elements. Factorization of the rotation matrix
with the PMNS matrix included explicitly in the case $3+3$ is
discussed in ref.~\cite{Xing:2011ur}. Here we give formulas for the
$3+1$ case.

The neutrino masses and their mixing angles are predicted from a given
neutrino mass matrix (the ``top-down'' method, as discussed
in~\cite{Dziewit:2011pd}). Exact analytical
expressions for the mixing angles, Dirac and Majorana phases, and
formulas for the non-physical phases can be given for the 3- and
4-dimensional cases. Only numerical solutions are possible in the
case of 2 or 3 additional neutrinos (i.e.\ 5-
or 6-dimensional~\cite{Xing:2011ur}
cases).

\bigskip
\noindent {\bf The 3-dimensional case}

First we parameterize the neutrino diagonalisation matrix by
including explicitly
the PMNS mixing matrix for the $3\times3$ mixing~\cite{Dziewit:2011pd}.
The neutrino mass matrix can be diagonalised
by a unitary transformation $U$, obtained
by the singular value decomposition method, 
see eq.~\refeq{Mtotal}.
Lets denote the complex
matrix elements in the following way:
\begin{equation}
\label{apAU}
U^{\mathrm (3\times3)}=
\left(
\begin{array}{ccc}
 x_1 & x_2 & x_3
  \\
 y_1 & y_2 & y_3
  \\
 z_1 & z_2 & z_3
\end{array}
\right).
\end{equation}
This matrix could be factorized into three terms
\begin{equation}
\label{apA2}
U^{\mathrm (3\times3)} =
  \hat{U}^{(3)}_{\phi} \cdot V_{\mr{PMNS}} \cdot \hat{U}^{(3)}_{\alpha},
\end{equation}
where $V_{\mr{PMNS}}$ is the standard PMNS mixing matrix~\cite{PDG2018} 
for Dirac neutrinos:
\begin{eqnarray}
\label{Upmns}
V_{\mr{PMNS}}&=&\left( \begin{array}{ccc}
1 & 0 & 0 \\
0 & \mr{c}_{23} & \mr{s}_{23} \\
0 & -\mr{s}_{23} & \mr{c}_{23}
\end{array} \right) \cdot
\left( \begin{array}{ccc}
\mr{c}_{13} & 0 & \hat{\mr{s}}_{13}^* \\
0 & 1 & 0 \\
-\hat{\mr{s}}_{13} & 0 & \mr{c}_{13}
\end{array} \right) \cdot
\left( \begin{array}{ccc}
\mr{c}_{12} & \mr{s}_{12} & 0 \\
-\mr{s}_{12} & \mr{c}_{12} & 0 \\
0 & 0 & 1
\end{array} \right) \notag \\
&=& \left(
\begin{array}{ccc}
 \mr{c}_{12} \mr{c}_{13} 
 & \mr{c}_{13} \mr{s}_{12} 
  & \hat{\mr{s}}_{13}^* \\
 -\mr{c}_{23} \mr{s}_{12} - \mr{c}_{12} \hat{\mr{s}}_{13} \mr{s}_{23}
 & \mr{c}_{12} \mr{c}_{23}- \mr{s}_{12} \hat{\mr{s}}_{13} \mr{s}_{23} 
  & \mr{c}_{13} \mr{s}_{23} \\
 \mr{s}_{12} \mr{s}_{23}-\mr{c}_{12} \mr{c}_{23}\hat{\mr{s}}_{13} 
 & -\mr{c}_{23} \mr{s}_{12} \hat{\mr{s}}_{13}-\mr{c}_{12} \mr{s}_{23} 
  & \mr{c}_{13} \mr{c}_{23}
\end{array}
\right) .
\end{eqnarray}
We used abreviations $\mr{c}_{ij} \equiv \cos\theta_{ij}$ and
$\hat{\mr{s}}_{ij} \equiv e^{i \delta_{ij}}\sin\theta_{ij}$, where
$\theta_{ij}$ and $\delta_{ij}$ are the rotation angle and the phase
angle, respectively.

The two diagonal phase matrices are defined as
\begin{eqnarray}
\hat{U}^{(3)}_{\phi} &=& \mr{diag} \left(e^{i \phi_1}, e^{i \phi_2}, e^{i \phi_3}
\right),
\\
\hat{U}^{(3)}_{\alpha} 
&=& \mr{diag} \left(1, e^{i \alpha_{21}/2}, e^{i \alpha_{31}/2} \right).
\end{eqnarray}
There are 9 parameters: 3~mixing angles $(\theta_{12}$,
$\theta_{13}$, $\theta_{23})$; 1~Dirac phase $\delta_{13} = \delta_{CP}$;
2~Majorana phases $\alpha_{21}$ and $\alpha_{31}$; and the matrix
$\hat{U}^{(3)}_{\phi}$ containing 3~non-physical and unmeasurable
phases $\phi_i$ ($i=1,2,3$).

Comparing eqs.~(\ref{apAU}) and (\ref{apA2}) we can find the relations
between the elements of the rotation matrix in a general form and its
parameters in the factorized form:
\begin{equation} \label{theta3x3}
\theta_{13}=\arcsin\left(\left|x_3\right|\right), 
\hspace{.2cm} 
\theta_{23}=\arctan\left(\frac{\left|y_3\right|}{\left|z_3\right|}\right), 
\hspace{.2cm} 
\theta_{12}=\arctan\left(\frac{\left|x_2\right|}{\left|x_1\right|}\right),
\end{equation}
\begin{equation} \label{DiracPh}
\delta_{13}
=\mr{arg}(x_2)-\mr{arg}(x_3)+\mr{arg}(y_3)
 -\mr{arg}\left(y_2 \left(1-|x_3|^2\right)+x_2 y_3 x_3^* \right),
\end{equation}
\begin{equation} \label{MajPh}
\frac{\alpha_{21}}{2}=\mr{arg}(x_2)-\mr{arg}(x_1), \quad 
\frac{\alpha_{31}}{2}= \mr{arg}(x_3)-\mr{arg}(x_1) +\delta_{13},
\end{equation}
\begin{equation} \label{phiPh1}
\phi_1=\mr{arg}(x_1), \quad \phi_2=\mr{arg}(x_1)-\mr{arg}(x_3)+\mr{arg}(y_3)-\delta_{13},
\end{equation}
\begin{equation} \label{phiPh2}
\phi_3=\mr{arg}(x_1)-\mr{arg}(x_3)+\mr{arg}(z_3)-\delta_{13}.
\end{equation}
These relations are obtained by comparing eq.~(\ref{apA2}) with the
corresponding
matrix elements from eq.~(\ref{apAU}) forming the upper-triangular matrix:
$x_1$, $x_2$, $x_3$, $y_2$, $y_3$,
and $z_3$. Other (numerically identical) solutions are possible, using
the diagonal elements and the matrix elements from the lower-triangular matrix
($y_1$, $z_1$, and $z_2$).

It should be noted that the Dirac phase can be evaluated using the Jarlskog 
invariant, for example expressed in the ``standard'' parameterization~\cite{PDG2018}
\begin{equation} \label{Jarskog}
J_{CP}=\mr{Im}\left( y_3 x_3^* x_2 y_2^*\right) 
= \frac{1}{8} \cos \theta_{13} \sin 2\theta_{12} \sin 2\theta_{23} \sin 2\theta_{13} \sin \delta_{13}.
\end{equation}
However, using this equation we need to be careful because $J_{CP}$ has the same 
value for $\sin (\delta_{13})$ and $\sin (\pi - \delta_{13})$, which gives a 
degeneracy of the $\delta_{13}$ values.

\bigskip
\noindent {\bf 4-dimensional case}

If there is one additional Majorana neutrino, decomposition of the
neutrino mass diagonalization matrix into factors including the PMNS
neutrino mixing matrix is more complicated.
Lets define the
2-dimensional rotation matrices in the 4-dimensional complex space,
similarly to ref.~\cite{Xing:2011ur},
\begin{equation*}
R^{(4)}_{12} =\left( \begin{array}{cccc} \mr{c}_{12} & \mr{s}_{12} & 0 & 0 \\
-\mr{s}_{12} & \mr{c}_{12} & 0 & 0 \\ 0 & 0 & 1 & 0 \\ 0 & 0 & 0 & 1 \end{array} \right), \qquad
R^{(4)}_{13} =\left( \begin{array}{cccc} \mr{c}_{13} & 0 & \hat{\mr{s}}_{13}^* & 0 \\ 0 & 1 & 0 & 0 \\
-\hat{\mr{s}}_{13} & 0 & \mr{c}_{13} & 0 \\ 0 & 0 & 0 & 1 \end{array} \right),
\end{equation*}
\begin{equation*}
R^{(4)}_{23} =\left( \begin{array}{cccc} 1 & 0 & 0 & 0 \\
0 & \mr{c}_{23} & \mr{s}_{23} & 0 \\ 0 & -\mr{s}_{23} & \mr{c}_{23} & 0 \\ 0 & 0 & 0 & 1 \end{array} \right), \qquad
R^{(4)}_{14} =\left( \begin{array}{cccc} \mr{c}_{14} & 0 & 0 &\hat{\mr{s}}_{14}^* \\
0 & 1 & 0 & 0 \\ 0 & 0 & 1 & 0 \\ -\hat{\mr{s}}_{14} & 0 & 0 & \mr{c}_{14} \end{array} \right),
\end{equation*}
\begin{equation}\label{apB1}
R^{(4)}_{24} =\left( \begin{array}{cccc} 1 & 0 & 0 & 0 \\
0 & \mr{c}_{24} & 0 & \hat{\mr{s}}_{24}^* \\ 0 & 0 & 1 & 0 \\ 0 & -\hat{\mr{s}}_{24} & 0 & \mr{c}_{24} \end{array} \right), \qquad
R^{(4)}_{34} =\left( \begin{array}{cccc} 1 & 0 & 0 & 0 \\
0 & 1 & 0 & 0 \\ 0 & 0 & \mr{c}_{34} & \hat{\mr{s}}_{34}^* \\ 0 & 0 & -\hat{\mr{s}}_{34} & \mr{c}_{34} \end{array} \right),
\end{equation}
and the phase matrices:
$\hat{U}^{(4)}_{\phi} = \mr{diag}
  \left(e^{i \phi_1},e^{i \phi_2},e^{i \phi_3},e^{i \phi_4}\right)$,
and
$\hat{U}^{(4)}_{\alpha} = \mr{diag}
  \left(1,e^{i \alpha_{21} /2},e^{i \alpha_{31}/2},1 \right)$.
Note that a shorter notation can be used to define the elements of
the rotation matrices:
\begin{equation}
\big[R^{(4)}_{jk}\big]_{a}^{\,\,b} =
\delta_{a}^{\,b} 
+ ( \mr{c}_{jk} - 1 ) 
  ( \delta_{a}^{\,j} \delta_{j}^{\,b} 
  + \delta_{a}^{\,k} \delta_{k}^{\,b} )
+ \hat{\mr{s}}_{jk}^{*} \delta_{a}^{\,j} \delta_{k}^{\,b} 
- \hat{\mr{s}}_{jk} \delta_{a}^{\,k} \delta_{j}^{\,b}\ ,
\end{equation}
where $\delta_{a}^{\,b}$ equals $1$, when $a=b$, or $0$, otherwise. This
notation is not restricted to the 4-dimensional case.

The unitary matrix $U^{\rm (4\times4)}$ is
parameterized by
\begin{equation}\label{apB3}
U^{\mr{(4\times4)}}=
\hat{U}^{(4)}_{\phi}\cdot
\left(R^{(4)}_{34} R^{(4)}_{24} R^{(4)}_{14}\right)\cdot
\left(R^{(4)}_{23} R^{(4)}_{13} R^{(4)}_{12}\right)\cdot
\hat{U}^{(4)}_{\alpha} ,
\end{equation}
with the PMNS matrix defined by a product of
three rotation matrices:
\begin{equation}\label{apB4} \left(
\begin{array}{cc} V_{\rm PMNS} & \mathbf{0} \\
\mathbf{0} & 1 \end{array} \right) =
\left(R^{(4)}_{23} R^{(4)}_{13} R^{(4)}_{12}\right),
\end{equation}
and the product of the other three rotation matrices $R_{i4}^{(4)}$
describes the mixing of the light neutrinos with the additional
heavy neutrino.
There are 16 parameters in this case, namely: 6 mixing angles
$(\theta_{12},\theta_{13},\theta_{23},
\theta_{14},\theta_{24},\theta_{34})$; 1~Dirac phase $\delta_{13}=\delta_{CP}$;
2~Majorana phases $\alpha_{21}$ and $\alpha_{31}$; 3~additional mixing
phases $\delta_{14},\delta_{24},\delta_{34}$; and 4~phases
$\phi_i$ ($i=1,2,3,4$).

For the model with $n_R=1$ the diagonalization matrix
$U$~(\ref{Mtotal}) is calculated numerically. Defining its elements as
\begin{equation}\label{apBU}
U^{\rm (4\times4)}=
\left(
\begin{array}{cccc}
 x_1 & x_2 & x_3 & x_4
  \\
 y_1 & y_2 & y_3 & y_4
  \\
 z_1 & z_2 & z_3 & z_4
  \\
 t_1 & t_2 & t_3 & t_4
\end{array}
\right)
\end{equation}
and comparing to eq.~(\ref{apB3}) we find the relations:
\begin{eqnarray}\label{apB5}
&& \theta_{12}= \arcsin \left(\frac{|x_2|}{\sqrt{b}}\right), \quad \theta_{13}= \arcsin \left(\frac{|x_3|}{\sqrt{a}}\right), \quad
\theta_{23}= \arcsin \left(\frac{|d|}{\sqrt{bc}}\right), \\
&& \theta_{14}= \arcsin \left(|x_4|\right), \quad
\theta_{24}= \arcsin \left(\frac{|y_4|}{\sqrt{a}}\right), \quad
\theta_{34}= \arcsin \left(\frac{|z_4|}{\sqrt{c}}\right),
\end{eqnarray}
\begin{eqnarray}\label{apB6}
&& \delta_{13}=\arg(x_2)-\arg(x_3)+\arg(d)-\arg\left( a\,b\,y_2 + b\,x_2\,y_4\, x^*_4 + d\,x_2\,x^*_3 \right),   \\
&& \delta_{14}=\phi_1-\arg(x_4), \quad
\delta_{24}=\phi_2-\arg(y_4), \quad
\delta_{34}=\phi_3-\arg(z_4), 
\end{eqnarray}
\begin{equation} \label{B7}
\frac{\alpha_{21}}{2}=\arg(x_2)-\arg(x_1), \quad \frac{\alpha_{31}}{2}=\arg(x_3) - \arg(x_1) + \delta_{13},
\end{equation}
\begin{eqnarray}\label{apB8}
&& \phi_1 = \arg(x_1), \\
&& \phi_{2}=\arg(x_1)-\arg(x_3)+\arg(d)-\delta_{13},   \\
&& \phi_{3}=\arg(x_1)-\arg(x_3)+\arg\left( a\,c\,z_3 + c\,x_3\,z_4\, x^*_4 + d\,z_4\,y^*_4 \right) - \delta_{13}, \\
&& \phi_4=\arg(t_4).
\end{eqnarray}
where:
\begin{eqnarray}
&& a=1-|x_4|^2, \qquad\qquad\quad\, b=1-|x_3|^2-|x_4|^2, \notag \\
&& c=1-|x_4|^2-|y_4|^2, \qquad d=a\, y_3 + x_3\, x^*_4\, y_4.
\end{eqnarray}

As $U^{(4\times4)}$ is unitary, there are relations between the elements. 
The expressions for the angles
do not contain all entries of the rotation matrix
$U^{(4\times4)}$, defined in eq.~\refeq{apBU}. 
The relations used
are obtained comparing eq.~(\ref{apB3}) with the matrix
elements from eq.~(\ref{apBU}) forming the upper-triangular matrix:
$x_1$, $x_2$, $x_3$, $x_4$, $y_2$, $y_3$, $y_4$, $z_3$, $z_4$,
and $t_4$. Other (numerically identical) solutions are possible
using the diagonal elements $x_1$, $y_2$, $z_3$, and $t_4$, and the matrix
elements $z_1$, $z_2$, $t_1$, $t_2$, and $t_3$.

\section{The two-Higgs-doublet model}
\label{Appendix-2HDM}

The most general 2HDM scalar potential of two doublets 
$\phi_1$ and $\phi_2$ is
\begin{eqnarray}
\label{pot2HDM}
V &=& m_{11}^{2} \phi_1^\dagger \phi_1 + m_{22}^{2} \phi_2^\dagger \phi_2
- \left( m_{12}^{2} \phi_1^\dagger \phi_2 + \mathrm{H.c.} \right) 
 \nonumber\\ & &
+\frac{\lambda_1}{2} \left( \phi_1^\dagger \phi_1 \right)^2
+ \frac{\lambda_2}{2} \left( \phi_2^\dagger \phi_2 \right)^2
+ \lambda_3\, \phi_1^\dagger \phi_1\, \phi_2^\dagger \phi_2
+ \lambda_4\, \phi_1^\dagger \phi_2\, \phi_2^\dagger \phi_1
 \nonumber\\ & &
+ \left[
  \frac{\lambda_5}{2} \left( \phi_1^\dagger \phi_2 \right)^2
  + \lambda_6\, \phi_1^\dagger \phi_1\, \phi_1^\dagger \phi_2
  + \lambda_7\, \phi_2^\dagger \phi_2\, \phi_1^\dagger \phi_2
  + \mathrm{H.c.}
  \right],
\end{eqnarray}
where the parameters $m_{11}^{2}$, $m_{22}^{2}$, and $\lambda_{1-4}$ 
are real numbers, whereas the remaining parameters $\lambda_{5}$,
$\lambda_{6}$, $\lambda_{7}$ and
$m_{12}^{2}$ in general can be complex. Since our main purpose 
of the paper is the analysis of the neutrino sector we restrict our 
analysis of the Higgs sector a CP-conserving Higgs potential with a 
softly broken $\mathbb{Z}_2$ symmetry, where $\lambda_5$ and $m_{12}^{2}$ 
are real, but $\lambda_6 = \lambda_7 = 0$.

We impose theoretical bounds on the potential, which allows us to
restrict the potential parameter space. Firstly, the unitarity
constraints set upper bounds on the parameters. These constraints come 
from the requirement that the scalar-scalar scattering amplitudes at 
tree-level must respect unitarity. Computation of the $S$ matrix 
for the scalar-scalar scattering amplitudes allows determination
of its eigenvalues 
\begin{eqnarray}
&& \Lambda_{1\pm} 
= \lambda_3 \pm \lambda_4, \quad \Lambda_{2\pm} 
= \lambda_3 \pm |\lambda_5|, \quad \Lambda_{3\pm} 
= \lambda_3 + 2\lambda_4 \pm 3|\lambda_5|
\enspace , \label{unCond1} \\
&& \Lambda_{4\pm} 
= \frac{1}{2}\left( 3\lambda_1 +3\lambda_2 
  \pm \sqrt{9(\lambda_1 - \lambda_2)^2 + 4(2\lambda_3 + \lambda_4)^2} \right)
\enspace , \\
&& \Lambda_{5\pm} 
= \frac{1}{2}\left( \lambda_1 +\lambda_2 
  \pm \sqrt{(\lambda_1 - \lambda_2)^2 + 4|\lambda_5|^2} \right)
\enspace ,   \\
&& \Lambda_{6\pm} 
= \frac{1}{2}\left( \lambda_1 +\lambda_2 
  \pm \sqrt{(\lambda_1 - \lambda_2)^2 + 4\lambda_4^2} \right)
\enspace . \label{unCond4}
\end{eqnarray}
Following ref.~\cite{Ginzburg:2005dt} we require that the 
eigenvalues~(\ref{unCond1})-(\ref{unCond4}) of all the scattering matrices 
should be smaller, in modulus, than $8\pi$, 
i.e.\ $\left|\Lambda_{i\pm}\right|<8\pi$.

To ensure a stable vacuum, the scalar potential has to be 
bound from below (BFB), i.e.\ there should be no direction in
the field space along which the potential tends to minus infinity. 
Necessary and sufficient conditions for the most general 2HDM scalar 
potential to be BFB were first derived in ref.~\cite{Maniatis:2006fs} 
and later in ref.~\cite{Ivanov:2015nea}. The procedure of 
ref.~\cite{Ivanov:2015nea} can only be handled numerically. 
For 2HDM scalar potentials, that are more constrained by symmetries,
like in potentials where one has $\lambda_6 = \lambda_7 = 0$, 
the necessary and sufficient BFB conditions can be
derived\footnote{A comprehensive derivation of the 
inequalities~(\ref{BFBCond}) is given in ref.~\cite{Jurciukonis:2018skr}.} 
as simple analytical expressions:
\begin{eqnarray}
&& \lambda_1 > 0, \quad \lambda_2 > 0, \notag 
\\
&& \lambda_3 > -\sqrt{\lambda_1 \lambda_2}, 
\quad \lambda_3 + \lambda_4 - |\lambda_5| > -\sqrt{\lambda_1 \lambda_2}
\enspace . 
\label{BFBCond}
\end{eqnarray}
We also apply the condition from ref.~\cite{Ivanov:2015nea}, which 
guarantees that the vacuum state has a lower value 
than all the other possible stability points of the potential:
\begin{equation}
\left[ \left( \frac{m^2_{H^{+}}}{v^2} + \frac{\lambda_4}{2}\right)^2
 - \frac{|\lambda_5|^2}{4}  \right]
\left[  \frac{m^2_{H^{+}}}{v^2}
 + \frac{\sqrt{\lambda_1 \lambda_2}-\lambda_3}{2}  \right] > 0
\enspace . \label{minCond}
\end{equation}

Finally we constrain the 2HDM scalar potential by applying the experimental 
bounds of the electroweak oblique parameters $S, T, U$. We require 
$S=0.02\pm0.10$, $T=0.07\pm0.12$, and $U=0.00\pm0.09$~\cite{PDG2018}. 
In our calculations we use expressions for the oblique parameters from 
ref.~\cite{Eriksson:2009ws} where they are determined in a convenient 
form for numerical calculations.

Our neutrino analysis requires a uniform coverage of the neutral
Higgs masses.
The numerical analysis uses Higgs masses
$m^2_h = \left(125.18\;\mathrm{GeV}\right)^2$ and 
$\{ m^2_H, m^2_A, m^2_{H^+} \} > m^2_h$ as input.
Studying the allowed ranges of the Higgs potential parameters,
we vary the angles
$\beta$ and $\beta-\alpha$ and the parameter $m_{12}^{2}$. 
The allowed range for $\alpha$ is fixed by the requirements 
$0\leq \beta \leq \pi/2$, and $-\pi/2\leq \beta - \alpha \leq \pi/2$. 
The Higgs masses $m^2_H$, $m^2_A$, and $m^2_{H^+}$ are varied up to 3~TeV.

Using the potential~(\ref{pot2HDM}), one arrives at the 
relations~\cite{Gunion:2002zf}
\begin{eqnarray}
\lambda_1 &=& \frac{1}{v^2 c^2_\beta} \left(m^2_h s^2_\alpha + m^2_H c^2_\alpha - m_{12}^{2} t_\beta \right), \\
\lambda_2 &=& \frac{1}{v^2 s^2_\beta} \left(m^2_h c^2_\alpha + m^2_H s^2_\alpha - m_{12}^{2} t^{-1}_\beta \right), \\
\lambda_3 &=& \frac{1}{v^2 s_\beta c_\beta} \left(\left(m^2_H-m^2_h\right) s_\alpha c_\alpha + 2m^2_{H^+} s_\beta c_\beta - m_{12}^{2} \right),   \\
\lambda_4 &=& \frac{1}{v^2 s_\beta c_\beta} \left(\left(m^2_A-2m^2_{H^+}\right) s_\beta c_\beta + m_{12}^{2} \right), \\
\lambda_5 &=& \frac{1}{v^2 s_\beta c_\beta} \left(m_{12}^{2} - m^2_A s_\beta c_\beta \right). \label{lambdas}
\end{eqnarray}
After computing the parameters $\lambda_{1-5}$ we validate the input
(i.e.\ the masses, angles, and $m_{12}^{2}$) by checking the
constraints described above hold. 

Following ref.~\cite{Haber:2006ue} for the CP-conserving limit 
we note that the quantities $b_i$ in 
table~\ref{table1} are related to the sign of the parameter $Z_6$
\begin{equation}
Z_6 = -\frac{1}{2} s_{2\beta} 
\left(\lambda_1 c^2_\beta - \lambda_2 s^2_\beta - (\lambda_3 + \lambda_4 + \lambda_5)c_{2\beta} \right)
\enspace . \label{Z6}
\end{equation}
However, in our case with $\lambda_6 = \lambda_7 = 0$, the sign of $Z_6$ 
anti-correlates with the sign of the angle $\beta-\alpha$: 
$\mathrm{sgn}(Z_6) = -\mathrm{sgn}(\beta-\alpha)$. Therefore the 
quantities $b_i$ in table~\ref{table1} depend only on the angle $\beta-\alpha$.



\bibliographystyle{JHEP}
\bibliography{gnm_v5}

\end{document}